\documentclass[aps,pra,twoside,twocolumn,10pt]{revtex4-2}
\usepackage[colorlinks=true, citecolor=blue, urlcolor=blue ]{hyperref}
\usepackage{epsfig,newlfont,amssymb,amsfonts,amsmath,bm,subfigure,palatino,mathtools,amsthm,soul,enumitem,color,graphics,graphicx,times,physics,xcolor}
\usepackage[normalem]{ulem}
\usepackage{multirow}
\usepackage{array}

\newcolumntype{P}[1]{>{\centering\arraybackslash}p{#1}}

\begin{document}
\title{
Disorder-induced quantum Fisher information in topological quantum systems
}
\author{Advay Burte$^{1,2}$, Keshav Das Agarwal$^{2,3}$, Leela Ganesh Chandra Lakkaraju$^{4,5}$,   Aditi Sen(De)$^{2,3}$}
\affiliation{$^1$ Department of Physics, BITS Pilani KK Birla Goa Campus, Zuarinagar 403726, Goa, India}
\affiliation{$^2$ Harish-Chandra Research Institute,  Chhatnag Road, Jhunsi, Prayagraj - 211019, India}
\affiliation{$^3$ Homi Bhabha National Institute, Training School Complex, Anushakti Nagar, Mumbai 400 094, India}
\affiliation{$^4$ Pitaevskii BEC Center, CNR-INO and Dipartimento di Fisica, Universit\`a di Trento, Via Sommarive 14, Trento, I-38123, Italy} 
\affiliation{$^5$ INFN-TIFPA, Trento Institute for Fundamental Physics and Applications, Via Sommarive 14, Trento, I-38123, Italy}

\begin{abstract}

The topological order of the Kitaev toric code is insensitive to disorder in its couplings since all star and plaquette operators commute; consequently, the stabilizer ground-state manifold is independent of the individual coupling strengths, and the phase remains stable against weak local perturbations. This insensitivity, however, is a property of the eigenstates and not of the dynamics they generate, and we show that the disorder sensitivity of a probe prepared outside that eigenbasis can be a metrological resource.
In particular, we investigate the estimation of the strength of three topological-order-breaking perturbations, the nonlinear Castelnovo-Chamon deformation, nearest-neighbor Ising interactions, and a magnetic field along horizontal edges, encoded on an initially separable probe through unitary evolution under the disordered toric code Hamiltonian. We demonstrate that the quenched average quantum Fisher information (QFI), a figure of merit for metrological precision, exceeds that of its ordered counterpart in suitable regimes of perturbation strength and encoding time. For the ordered case, we further identify optimal product probes, for which the quadratic and cubic terms of the short-time expansion vanish identically, leaving a transient quartic scaling of QFI. Interestingly, this window persists as long as the multipartite entanglement generated by the encoding continues to increase, and it survives under local dephasing, bit-flip, and amplitude damping noise acting independently on each site, with the non-unital channel being the least detrimental. 

\end{abstract}

\maketitle

\section{Introduction}
\label{sec:intro}

Topologically ordered phases are characterized by non-local properties of their ground states~\cite{Wen1995, Wen2002, Kitaev03}, in contrast to conventional phases of matter, which are described by local order parameters and symmetry breaking within the Landau paradigm. These include a topology-dependent ground-state degeneracy and long-range entanglement, whose universal contribution can be captured by the topological entanglement entropy~\cite{Wen2006, kitaev_preskill_2006}. Among the paradigmatic models developed to realize topological order, the Kitaev toric code is particularly prominent due to its exact solvability and the stabilizer structure~\cite{Kitaev03}. Its locally indistinguishable ground states and stability under sufficiently weak local perturbations provide a natural setting for storing quantum information~\cite{Bravyi2010}, and have motivated the surface-code framework for quantum error correction~\cite{kitaev_preskill_2002}. Strong perturbations can, however, destroy the topological order, allowing the toric code to serve as a paradigmatic model for investigating topological quantum phase transitions~\cite{CastelnovoChamon08,Karimipour_2013,Halasz_2012}. It is important to note that all these properties concern the eigenstates of the Hamiltonian, and not the dynamics that they generate. A probe prepared outside the stabilizer eigenbasis evolves in a manifestly coupling-dependent manner, so that coupling disorder, invisible in equilibrium, suppresses the recurrences of the subsequent unitary evolution~\cite{RahmaniChamon10}, while the fate of the topological order itself after a quench depends on the quench performed~\cite{Tsomokos09, Zeng16, HalaszHamma13}. Whether this disorder sensitivity of the dynamics, rather than the robustness of the eigenstates, can serve as a metrological resource is the question we address here.

\begin{figure}
    \centering
    \includegraphics[width=\columnwidth]{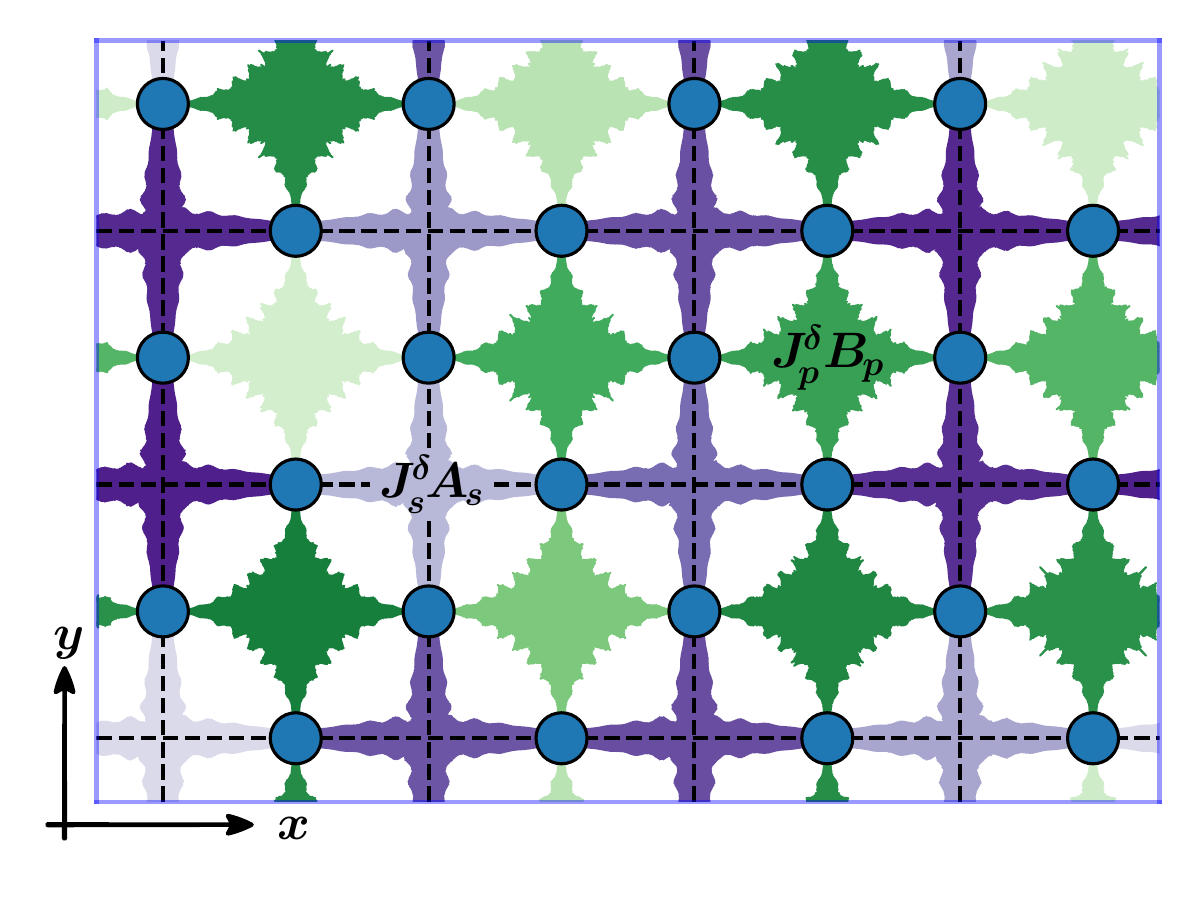 }
    \caption{The disordered Kitaev toric code model on the two dimensional lattice (dashed lines) with the qubits (blue circles) on the edges and periodic boundary conditions. The qubits near a vertex $s$ of the lattice form a star $A_s$ (purples), and the qubits on the edges of square $p$ form a plaquette \(B_p\) (greens), with disordered (site-dependent) interaction strength $J_s^{\delta}$ and $J_p^{\delta}$ where $\delta$ denotes the disorder strength. The strength of perturbations are dynamically encoded on initially product state in the presence of disorder.
    }
    \label{fig:toric_code_lattice}
\end{figure}
Quantum sensing typically exploits quantum coherence and correlations to estimate physical quantities such as magnetic fields, temperatures, and coupling strengths with enhanced precision~\cite{GLM04, GLM06, Sensing_RMP_2017, PezzeRMP18} in a static scenario. The sensitivity of a quantum probe to an encoded parameter is quantified by the quantum Fisher information (QFI), which bounds the achievable precision through the quantum Cram\'{e}r-Rao bound~\cite{Helstrom1976,braunstein1994,Paris09}. In a dynamical sensing protocol, the parameter is instead imprinted on an initially prepared state through evolution under a parameter-dependent Hamiltonian~\cite{BoixoPRL07,BoixoDynamics08,giovannetti_nature, PangBrun14}. Such protocols have been investigated using periodic driving in spin systems~\cite{utkarsh2021} and Floquet time crystals~\cite{IeminiTimeCrystal24}, highlighting the role of many-body dynamics in sensing performance. The resulting QFI depends on the initial probe state, the encoding time, and the interplay among the different terms of the Hamiltonian~\cite{PangBrun14}. 
This naturally raises the question of how the stabilizer interactions of a topologically ordered model shape the sensitivity of a probe when an unknown parameter is encoded dynamically, starting from a product state that itself carries no topological order. Here, we answer this by investigating the precision of estimating perturbations to the toric code encoded during the evolution, focusing on the temporal scaling of the QFI, its dependence on the initial probe state, its relation to entanglement generated during the encoding dynamics, and its robustness against preparation imperfections and noise.

Experimental realizations of many-body systems are naturally affected by imperfections and inhomogeneities, which can be modeled as disorder in the system parameters~\cite{Fort2005,Roati2008,Fallani_2007,White_2009}. Interestingly, it has been shown that in various systems, quenched disorder~\cite{Sherrington_1975, Edwards1975,  Derrida_1980, Aizenman_1989, Rieger_1994, Ahufinger_2005, Lewenstein2007, Shapiro_2012, Vojta2019} can play a constructive role in enhancing system properties, such as magnetization~\cite{Niederberger_2008, Niederberger2009, Niederberger_2010, SanchezPalencia2010, Bera_2016},  quantum correlations~\cite{Mishra2016,Sadhukhan_2016}, and in quantum devices like the performance of quantum thermal devices~\cite{Ghosh_2020,Konar_2022}, and the precision of quantum sensors~\cite{Yousefjani2023,He_2023,Sahoo_2024,Bhattacharyya_2024,Mothsara_2025,Sarkar_2025,Pezze_2026,Vishnupriya_2026}. These findings motivate us to examine both the robustness and the potential for disorder-induced enhancement in the dynamical encoding of perturbation strengths in the toric code. Moreover, unavoidable interactions with the environment can degrade quantum resources, such as coherence and correlations, that are essential for quantum information processing\cite{open_quan_book, Zurek2003, Sen_2006} and quantum sensing~\cite{Wasilewski_2010, Escher2011, DemkowiczDobrzaski2012, Koodyski2013, Alipour_2014, haase2016_open_review, Falaye2017}. It is therefore important to investigate how local noise affects the evolving probe and, consequently, the noise tolerance of a sensing protocol initiated from a product state.

To assess the robustness of the topological models in  sensing,  we estimate the strength of three perturbations dynamically encoded in an optimal product probe state to the toric code, the nonlinear Castelnovo-Chamon deformation, nearest-neighbor Ising interactions, and a magnetic field acting on spins along horizontal edges. We then introduce Gaussian-distributed disorder into the star or plaquette interaction terms of each model to examine how disorder affects metrological performance (see Fig.~\ref{fig:toric_code_lattice}). We report that disorder can amplify the quenched average QFI in suitable parameter regimes, with the response depending on the perturbation and the encoding time -- we refer to this phenomenon as {\it disorder-induced QFI}. In the absence of disorder,  we identify product probes for which the QFI exhibits transient quartic growth, 
\(\sim t^4\), and  show that the duration of this regime is connected with the increase of multipartite entanglement generated by the encoding dynamics. We further observe that the transient QFI scaling persists even under local dephasing, bit-flip, and amplitude-damping noise before decaying at longer times, with the QFI being most robust to amplitude damping among the three noise channels.

The paper is organized as follows. In Sec.~\ref{sec:background}, we introduce the quantum metrological framework, the toric code, and the three perturbations under consideration. Section~\ref{sec:disorderQFI} presents the effect of disorder on the dynamical QFI. In Sec.~\ref{sec:dynamics}, we emphasize the condition for transient quartic growth and investigate its connection with dynamically generated multipartite entanglement. Section~\ref{sec:noise} examines the effect of local noise, and Sec.~\ref{sec:conclusion} summarizes our findings.

\section{Quantum precision limits and the toric code Hamiltonian}
\label{sec:background}
 
We begin by introducing the central figure of merit that quantifies the performance of quantum sensors, namely the quantum Fisher information~\cite{braunstein1994,Paris09}. The QFI of a quantum state with respect to the encoded parameter measures the sensitivity of the quantum state under the encoding and provides an upper bound to the precision for an arbitrary measurement scheme used for parameter estimation. We then discuss the Kitaev toric code Hamiltonian, as well as the various perturbations that are encoded dynamically onto a product pure state. The principal objective is to estimate the corresponding perturbation strength, thereby providing the framework for assessing the sensing capabilities of the perturbed Kitaev toric code in the dynamical regimes.

\subsection{Quantum metrology}

The central task of quantum metrology is to estimate an unknown parameter $\theta$ encoded in a quantum state $\rho_\theta$. In a dynamical sensing protocol, a quantum probe state $\rho(t=0)$ evolves under a $\theta$-dependent process, $\mathcal{E}_{\theta,t}$, which imprints the information about $\theta$, leading to the state $\rho_\theta(t) = \mathcal{E}_{\theta,t}(\rho_0)$. The efficiency of encoding can be obtained by calculating the QFI as
$  \mathcal{F}_{\theta}[\rho_\theta] =\operatorname{Tr}\!\left[\rho_\theta\,L_\theta^2\right], \
    \partial_\theta\rho_\theta = \frac{1}{2}\!\left(L_\theta\,\rho_\theta + \rho_\theta\,L_\theta\right),
$ where $\partial_\theta\ \equiv \frac{\partial}{\partial\theta}$, and  $L_\theta$ is the symmetric logarithmic derivative (SLD) with respect to the parameter $\theta$, which is also the (Hermitian) operator providing the optimal measurement strategy. Writing $\rho_\theta = \sum_{j=1}^d\lambda_j \ket{\lambda_j}\bra{\lambda_j}$ in its spectral decomposition, the QFI can be computed as~\cite{LiuCTP14, LiuJPA20}
\begin{equation}
    \mathcal{F}_\theta[\rho_\theta] = \sum_{j,k=1}^d \frac{2 \lvert\bra{\lambda_j}\partial_\theta\rho_\theta\ket{\lambda_k}\rvert^2}{\lambda_j + \lambda_k}.
    \label{eq:mixed_QFI_expr}
\end{equation}
It provides a lower bound on the variance in the estimate of the parameter $(\Delta \theta)^2 = \langle \theta^2\rangle - \langle\theta\rangle^2$, which can obtained via the measurement outcomes with any locally unbiased estimator as $(\Delta \theta)^2\;\geq\; (M\,\mathcal{F}_\theta[\rho_\theta])^{-1}$ with $M$ being the number of independent measurement trials, known as the quantum Cram\'{e}r-Rao bound (QCRB)~\cite{Helstrom1976, Holevo1982, braunstein1994}. Note that for a pure state $|\psi_\theta\rangle$, the above expression leads to 
\begin{equation}
     \mathcal{F}_\theta[\ket{\psi_\theta}]= 4\left(
        \braket{\partial_\theta\psi_\theta}{\partial_\theta\psi_\theta} - \left|\braket{\psi_\theta}{\partial_\theta\psi_\theta}\right|^2 \right). \label{eq:QFI_expr}
\end{equation}
As the bound is attainable for single-parameter estimation, a higher QFI implies a smaller achievable variance, i.e., a higher precision, and, therefore, the state $\rho_\theta$ carries more information about the parameter $\theta$.

\subsection{Toric code and the perturbations encoded}

We investigate dynamical quantum sensing in the Kitaev toric code model defined on a two-dimensional ($2$D) rectangular lattice comprising $L_x$ and $L_y$ blocks along the $x-$ and $y-$ directions, respectively. The system consists of $N=2L_xL_y$ spin-$1/2$ particles (qubits) residing on the edges of the $L_x\times L_y$ lattice with periodic boundary conditions~\cite{Kitaev03}. 
The set of four qubits around each vertex of the lattice forms a star $s$, and similarly, the set of four qubits on the edges of each block is termed as a plaquette $p$ (see Fig. \ref{fig:toric_code_lattice} for a schematic). The Kitaev toric code Hamiltonian can be represented as
\begin{equation}
    H^{\delta}_{TC} = -\!\left(\sum\nolimits_s J_s^{\delta} A_s +\sum\nolimits_p J_p^{\delta} B_p \right)    \label{eq:Hamil_TC}
\end{equation}
where $ A_s= \prod_{j\in s} \sigma_j^x $, $ B_p = \prod_{k\in p} \sigma_k^z,
$ and $ \sigma^{\alpha}_j (\alpha=x,y,z)$ are the Pauli matrices $\left(\sigma^z_j=\ket{\uparrow}\!\bra{\uparrow}-\ket{\downarrow}\!\bra{\downarrow}\right)$ on the $j$-th qubit. Here, $J_s^\delta$ ($J_p^\delta$) are the site-dependent interaction strengths at the vertex $s$ (plaquette $p$), which can result from lattice defects or disorder in the interactions with the disorder strength $\delta$ (see Fig. \ref{fig:toric_code_lattice}). Specifically, we sample $J_p^\delta$ and $J_s^\delta$ from Gaussian distributions with corresponding standard deviations \(\delta_s\) and $\delta_p$ as mentioned in the succeeding sections.
We refer to the model as the disordered Kitaev toric code model, which, with vanishing $\delta_s$ and $\delta_p$, represents the disorder-free case.  Specifically, when all interaction strengths are equal, i.e., $J_s^\delta = J_p^\delta = J$, the model can simply be called the Kitaev toric code, denoted as $H_{TC}$. The ground states (lowest energy eigenstates for $J>0$) of the $H_{TC}$ are four-fold degenerate and possess intrinsic topological order~\cite{Kitaev03, Bravyi2010}, which aids in quantum error correction~\cite{kitaev_preskill_2002}.

While the intrinsic topological order is robust to weak perturbations, it is broken in presence of strong perturbations, giving a topological quantum phase transition point at finite perturbation strengths~\cite{TrebstToric07, CastelnovoChamon08, Karimipour_2013, Halasz_2012, santra2014, Zhang_2022}. In this work, we encode the perturbation amplitude of the Kitaev toric code model with and without disorder, on a quantum state. Specifically, the perturbed system Hamiltonian with disorder reads as
\begin{align}
    \label{eq:TC0_per}
    H^{\delta}_{TC}(\theta) = H^{\delta}_{TC} + J H_X(\theta),
\end{align}
where $H_X(\theta)$ denotes the perturbations of type $X$ with the strength $\theta$. In this work, our goal is twofold -- first to study whether topological properties can provide a benefit in quantum metrology; and second, to determine whether disorder in topological systems retains or enhances metrological precision.  

\emph{Castelnovo-Chamon perturbation}.
The nonlinear perturbation studied by Castelnovo and Chamon~\cite{CastelnovoChamon08}, is the $z$-deformation on the stars given by
\begin{equation}
    H_{\mathrm{CC}}(\theta) = \sum_{s} \prod_{j\in s} e^{-\theta\hat{\sigma}_j^{z}}.
    \label{eq:dis_cc}
\end{equation}
with $X\equiv \mathrm{CC}$. Note that \(\theta = 0\) gives $H_{\mathrm{CC}}(0)=\mathbb{I}$, denoting no perturbation, reducing to the original Kitaev toric code. In the \(\theta \to \infty\) limit, the ground states are fully polarized in the $z$-direction. This indicates that as \(\theta\) increases from \(0\) to \(\infty\), the system undergoes a topological phase transition, i.e., it crosses from a topological phase to a trivial phase, occurring at \(\theta_c= \frac{1}{2}\ln(1+\sqrt{2})\sim 0.441\)~\cite{CastelnovoChamon08}.\\

\emph{Ising perturbation}. 
Another prominent perturbation is the nearest-neighbor Ising-type interactions on the $2$-D lattice~\cite{Karimipour_2013},
\begin{equation}
    H_{\text{Ising}} (\theta) = - \theta \sum_{\langle j,k\rangle}  \ \sigma_j^x \sigma_{k}^x,
    \label{eq:dis_is}
\end{equation}
where $\langle j,k\rangle$ denotes the pair of nearest-neighbor sites on the toric code lattice and $X\equiv \text{Ising}$. The corresponding Kitaev toric model $H_{TC}(\theta)$ showcases a phase transition between a topologically ordered phase and an antiferromagnetic phase at \(\theta_c  \sim 1/6\)~\cite{Karimipour_2013}.\\

\emph{Magnetic field on spins along horizontal edges}. 
This corresponds to the $z$-direction magnetic field only on the horizontal edges of the lattice~\cite{Halasz_2012, Zhang_2022}, resulting in
\begin{equation}
    H_{\text{Mag}} (\theta) = -\theta \sum_{k\in \mathrm{E_x}}\sigma_k^z,
    \label{eq:dis_mag}
\end{equation}
where $X\equiv \text{Mag}$, and $\mathrm{E_x}$ are the set of qubits only on edges in the $x-$direction. The corresponding perturbed Kitaev toric model $H_{TC}(\theta)$ has a phase transition at \(\theta_c= 1\)~\cite{Halasz_2012, Zhang_2022}, at which point a transition from a topologically ordered phase to a paramagnetic phase occurs.

\section{Disorder-induced QFI}
\label{sec:disorderQFI}

\begin{figure*}
    \centering
    \includegraphics[width=1\linewidth]{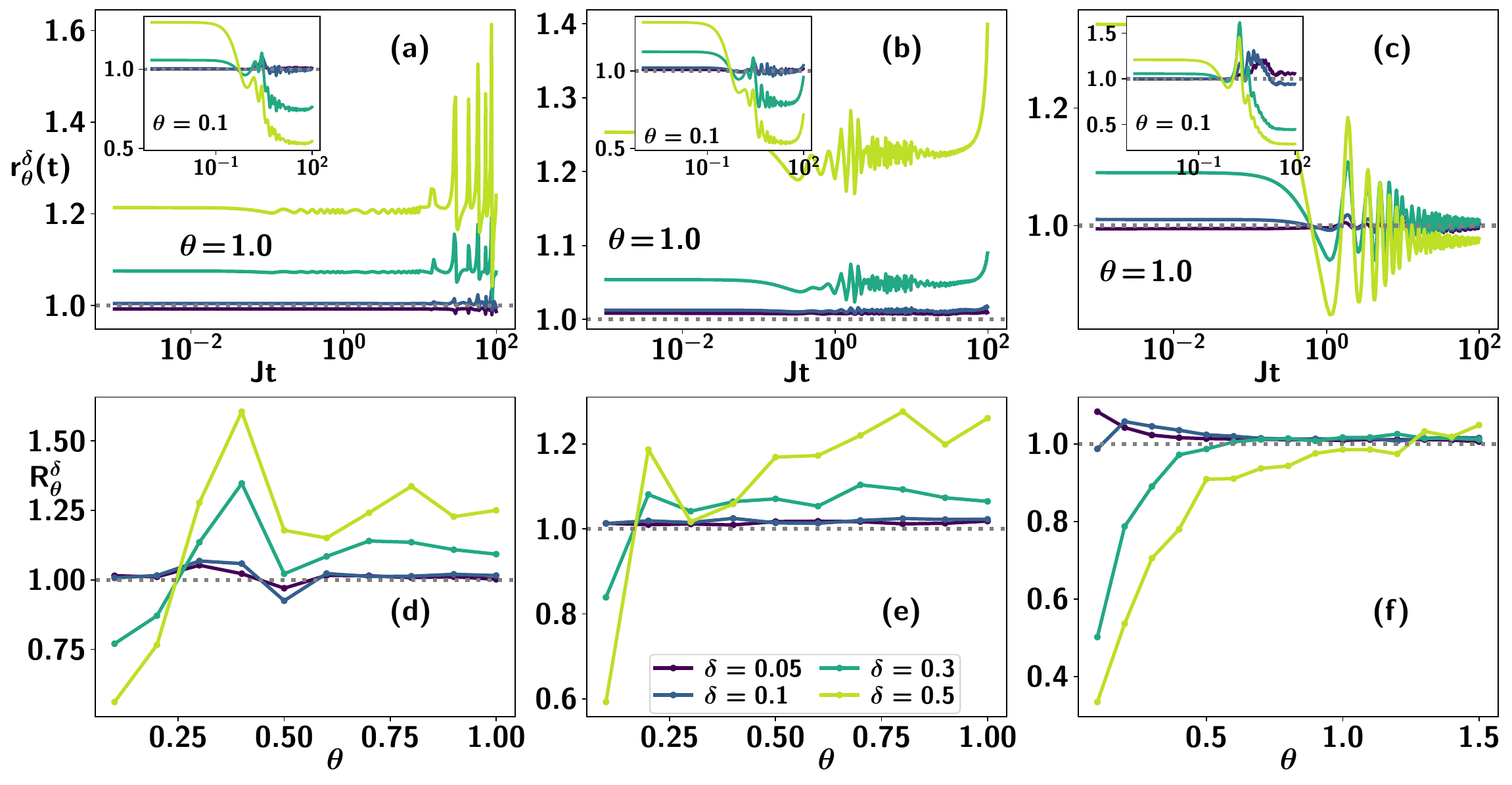}
    \caption{{\bf Disorder-enhanced dynamical quantum sensing of the perturbations on the toric code.} (a)-(c) $r^{\delta}_{\theta}(t)$ (ordinate), the ratio between the disorder-averaged QFI and  the QFI of the corresponding ordered systems, against the encoding time $Jt$ (abscissa) for perturbation $\theta=1.0$ ($\theta=0.1$ in the insets). The bottom panels [(d)-(f)] show the time-averaged ratio $R^{\delta}_{\theta}$ (ordinate) against the encoded perturbation $\theta$ (abscissa) for various disorder strengths $\delta$. Here (a),(d) represent Castelnovo-Chamon, (b),(e) are for Ising, and (c),(f) for magnetic field perturbations. Both the quantities in the ordinate are defined in Eq.~(\ref{eq:disorder_ratio_and_average}), and the horizontal dotted line marks unity, above (below) which the disorder amplifies (degrades) the QFI. Darker to lighter shades represent increasing disorder strength, $\delta = 0.05, 0.1, 0.3$ and $0.5$. For the Castelnovo-Chamon and the Ising perturbations, the QFI is enhanced with increasing $\delta$ relative to the ordered system in the topologically trivial phase, whereas in the topologically ordered phase, an initial window of enhancement gives way to a degradation that grows with $\delta$. For the magnetic field perturbations, such disorder-induced enhancement is seen for high values of $\theta$. The system size is $N=12$, the quenching is performed over \(200\) realizations and the time scale for averaging is $Jt \leq 100$. All the axes are dimensionless.}
    \label{fig:disorder_qfi}
\end{figure*}

The encoding is performed by the unitary (closed) evolution of the initial state $|\psi(0)\rangle$ governed by the disordered perturbed toric code Hamiltonian $H^{\delta}_{TC} (\theta)$, resulting in the time-evolved state as
\begin{align}
    \ket{\psi^{\delta}_{\theta}(t)} &= \exp\Big[-i H^{\delta}_{TC}(\theta) t\Big] |\psi(0)\rangle, 
    \label{eq:TC_dis}
\end{align}
for all three perturbations $\mathrm{CC}$, $\mathrm{Ising}$ and $\mathrm{Mag}$ field respectively. For each perturbation, the initial state is taken as a separable eigenstate of the perturbation Hamiltonian $H_X(\theta)$, such that any non-trivial encoding of $\theta$ in the time-evolved state $\ket{\psi^{\delta}_{\theta}(t)}$ occurs due to the presence of toric code interactions. Specifically, for the CC and magnetic field perturbations, we take $\ket{\psi(0)} = \ket{\uparrow}^{\otimes N}$. In contrast, for the Ising case, we take  $\ket{\psi(0)} = \ket{+}^{\otimes N}$, where $\ket{+}_j = \frac{\ket{\uparrow}+\ket{\downarrow}}{\sqrt{2}}$, is the  $+1$ eigenstate of $\sigma_j^x$. One can easily verify that starting from an eigenstate with a different eigenvalue yields the same QFI. We select this set of initial states after optimizing over arbitrary pure product states for quartic scaling behavior of QFI with time, which we discuss in the succeeding section.

In our study, the evolution operator involves random couplings, and hence quenched averaging has to be performed to determine the metrological performance \cite{Bhattacharyya_2024,Mothsara_2025,Sarkar_2025,Pezze_2026,Vishnupriya_2026}. In particular, we compute the quenched average QFI for the perturbation parameter $\theta$ as
\begin{align}
    \langle\mathcal{F}^{\delta}_\theta(t)\rangle \!&=\! \int P_{s}(J_s^\delta) P_{p}(J_p^\delta) \mathcal{F}_\theta\left[\ket{\psi^{\delta}_{\theta}(t)}\right] \ d J_s^\delta d J_p^\delta, \nonumber \\
    P_{a=s,p}(Y) &= \frac{1}{\delta_a\sqrt{2\pi}}\exp\left(\frac{-(Y-J)^2}{2\delta_a^2}\right),
\end{align}
where $\mathcal{F}_\theta\left[\ket{\psi_\theta}\right]$ is defined in Eq.~\eqref{eq:QFI_expr}, $P_{a}$ ($a=s,p$) are the probability distributions from which $J_a^\delta$ are randomly chosen, and $\delta = 0$ represents the ordered case for which the QFI at time $t$ is denoted as $\mathcal{F}_\theta(t)$. Note that the quenched averaging is performed under the assumption that the observation time is much shorter than the time taken for the disordered parameters $J_s^\delta$ and $J_p^\delta$ to equilibrate, which implies that the disordered parameters are frozen during the dynamics~\cite{SanchezPalencia2010, Bera_2016, Mishra2016, Sadhukhan_2016}. Here, we ask the following question: can the magnitude of the QFI for the perturbed parameter be higher in the disordered case than that of the ordered one, at least in some regimes? Importantly, can we get this disorder-induced QFI, if it exists, for all kinds of perturbations? Let us answer this by considering all the perturbations in Eqs. (\ref{eq:dis_cc}) - (\ref{eq:dis_mag}).

Before stating the results, let us first state a few observations and the figures of merit that we study based on the quenched average QFI. Firstly, the $\mathrm{(CC)}$ and magnetic field  perturbations commute with the interactions $B_p$ on the plaquettes $p$, i.e., $[H_{X}(\theta), B_p]=0$. Therefore, the effect of disorder on plaquette interactions results only in an overall global phase difference of the time-evolved state, and the corresponding QFI is insensitive to the disorder on the plaquettes. It immediately implies that $\langle \mathcal{F}_{\theta}^{\delta}(t) \rangle$ reduces to $\mathcal{F}_\theta (t)$ when randomness is present only in the plaquettes and not in the stars, having $\delta_s = 0$ with $\delta_p \neq 0$, while this is not the case when $J_s^\delta$ are randomly chosen. Secondly, the opposite is true for the $\mathrm{Ising}$ perturbation, i.e., $H_{\mathrm{Ising}}(\theta)$ commutes with the star terms $[H_{\mathrm{Ising}}(\theta),A_s]=0$, and the disorder on the plaquette affects the corresponding QFI. Therefore, we restrict our attention to the relevant disorder type for each model, i.e., disorder on the star for $X \in \{\mathrm{CC}, \mathrm{Mag}\}$ and on the plaquette for $X \equiv \mathrm{Ising}$, and denote the relevant disorder strength as $\delta$ and drop the subscripts $s/p$ for brevity.

To quantify the effects of disorder on the QFI, we introduce two quantities: (1) the disorder-induced  QFI gain, defined as the ratio of the quenched average QFI to the QFI in the absence of disorder; and (2) the time-averaged disorder-induced QFI gain, obtained by averaging the former quantity over time to eliminate its time dependence. Mathematically, they are given respectively as
\begin{equation}
    r^{\delta}_\theta(t) = \frac{\langle\mathcal{F}^{\delta}_\theta(t)\rangle}{\mathcal{F}_\theta(t)} ~~ \text{and}~ \ R^{\delta}_\theta=\frac{1}{T}\int_0^T r^{\delta}_\theta(t) dt.
    \label{eq:disorder_ratio_and_average}
\end{equation}
$R^{\delta}$ quantifies the overall amplification ($R^{\delta}_\theta>1$) or reduction ($R^{\delta}_\theta<1$) of the QFI by the disorder over a time period $T$~\cite{Vishnupriya_2026}. We now examine how disorder affects the encoding of each perturbation strength in its corresponding initial product state.

\emph{Castelnovo-Chamon perturbation.} At low perturbation strength $\theta$, 
the disorder in star interactions causes an enhancement in the disorder-averaged QFI over the no-disorder QFI at initial times, i.e., $r^{\delta}_\theta(t)>1$  when $t\lesssim 1$. In particular, the amplification increases with increasing disorder-strength $\delta$, i.e., $r^{\delta}_\theta(t\lesssim1) > r^{\tilde{\delta}}_\theta(t\lesssim1)$ when $\delta>\tilde{\delta}$. However, the enhancement fades with the increase of time and eventually there is a reduction in the disorder-averaged QFI with respect to the QFI in the ordered case, leading to $r^{\delta}_\theta(t\gg1)<1$, which decreases with increasing $\delta$. 
Interestingly, at higher values of $\theta$, 
the quenched average QFI is amplified in both the initial and the large times, i.e, $r^{\delta}_\theta(t)>1$ for large $\theta$, which increases with higher disorder strength, as illustrated in Fig.~\ref{fig:disorder_qfi}(a), for $\theta = 0.1$ (topologically non-trivial) and $\theta=1.0$ (topologically trivial) phases. For fixed system size and choosing $T=100$, we observe that the effect of disorder from the degradation to amplification changes at $\theta\sim 0.25$. Specifically,  with $R^{\delta}_{\theta\lesssim0.2}<1$ while $R^{\delta}_{\theta\gtrsim0.2}>1$, and these effects increase with disorder strength (see Fig.~\ref{fig:disorder_qfi}(d)), thereby establishing disorder-induced enhancement in the topologically trivial phase.

\emph{Ising perturbation.} When the toric code encodes the Ising perturbation on the initial product state, the disorder amplifies the corresponding (averaged) QFI $\langle\mathcal{F}_\theta^{\delta}(t)\rangle$ at large times $t\gg1$. At the same time, it reduces QFI, i.e., $r_\theta^{\delta}(t)<1$ at small times $t\ll1$. On increasing disorder strength $\delta$, $r^{\delta}_\theta(t)$ continues to be greater than unity for all values of $\theta$ when $t \gg 1$ (see Fig. \ref{fig:disorder_qfi} (b)). This disorder-induced QFI can also be captured by analysing $R_\theta^{\delta}$, which turns out to be above unity for high values of $\theta$ (see Fig. \ref{fig:disorder_qfi} (e)), although low $\theta$ values cannot provide any benefit due to disorder.

\emph{Magnetic field on spins along horizontal edges.} 
In the dynamical encoding of the magnetic field via the toric code, the quenched average QFI gets enhanced for small times, irrespective of the values of $\theta$, while the advantage is more pronounced for large values of $t$ with large values of $\theta$. Precisely, $R_\theta^{\delta}>1$ for large values of $\theta$, thereby confirming disorder-induced gain. On the other hand, $r_\theta^{\delta}>1$ in the transient regime for all $\theta$ (see Fig. \ref{fig:disorder_qfi} (c),(f)). This disorder-induced metrological gain remains evident with the low to moderate disorder strength.  

Taken together, the three perturbations lead to a picture that is more intricate than one would anticipate. A natural expectation is that the disorder cannot be beneficial deep inside the topological phase, where the robustness of the topological order protects the model against weak local perturbations, and that the benefit should set in once the topological order is destroyed. None of the three perturbations follows this expectation. For the $\mathrm{CC}$ deformation, the time-averaged gain crosses unity at $\theta \sim 0.2$, already inside the topologically ordered phase, $\theta_{c} \simeq 0.441$. For the $\mathrm{Ising}$ perturbation, the entire range displayed in Fig.~\ref{fig:disorder_qfi}(e) lies above $\theta_{c} \sim 1/6$, yet the crossover occurs only at $\theta \sim 0.2$, well inside the trivial phase. The magnetic perturbation is displaced in the same direction but much further, with the gain appearing only for $\theta \gtrsim 1.25$ against $\theta_{c}=1$, and remaining of the order of a few percent, in contrast to the amplification by factors of about $\sim 1.5$ and $\sim 1.2$ for the $\mathrm{CC}$ and the $\mathrm{Ising}$ perturbations, respectively for $\delta=0.5$ and system-size $N=12$. Hence, within the finite systems studied here, the loss of topological order is neither necessary nor sufficient for a disorder-induced metrological gain, and neither the location of the crossover nor the magnitude of the gain is controlled by the equilibrium topological quantum phase transition. Identifying the mechanism that decides which perturbations benefit from the disorder requires further investigation.

\section{Enhanced scaling of QFI in dynamical metrology}
\label{sec:dynamics}

Let us now exhibit that the dynamical encoding of the perturbation parameter of a topological system on the quantum probe can provide an advantage over classical encoding strategies, given by the scaling of QFI with time. In classical dynamical encoding, the QFI increases quadratically in time, known as the standard quantum limit (SQL), whereas super-quadratic temporal scaling of QFI is shown to be obtained in quantum systems at transient times for optimal initial probes~\cite{rams2018, Puig2025}. In our case, the unitary evolution of a quantum probe via the perturbed Kitaev toric code Hamiltonian $H_{TC}(\theta)$ with site-independent interactions and without disorder, encodes the interaction parameter $\theta$ on the quantum state. We aim to determine the scaling of the QFI with the resources, i.e., with time and system size, which illustrates the efficiency of the dynamical encoding and is central to designing good sensing protocols.
\begin{figure*}
    \centering
    \includegraphics[width=1\linewidth]{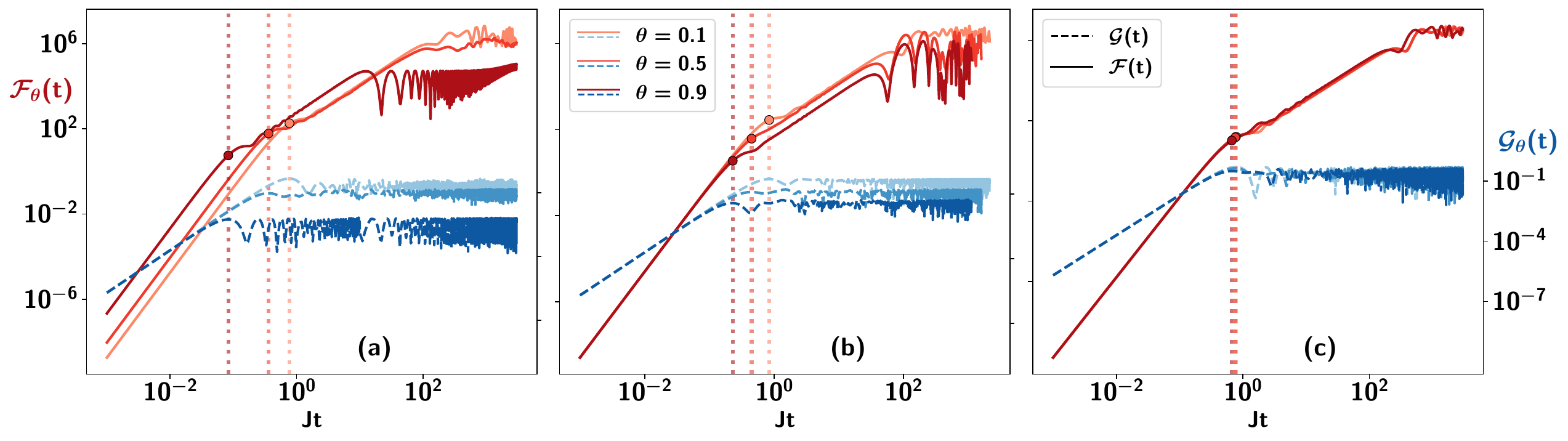}
    \caption{{\bf Multipartite entanglement growth during the quartic scaling of QFI in transient time.} The QFI $\mathcal{F}_\theta(t)$ (left ordinate, solid red curves) and the multipartite entanglement measure, GGM, $\mathcal{G}_\theta(t)$ (right ordinate, dashed blue curves), against time $t$ (abscissa) for (a) Castelnovo-Chamon, (b) Ising, and (c) magnetic field perturbations on horizontal edges of the toric code, with $N=20$ and interaction strengths, $J_s =J_p =J$. The initial probes are $\ket{\uparrow}^{\otimes N}$ for (a) and (c), and $\ket{+}^{\otimes N}$ for (b). Lighter to darker shades represent increasing perturbation strengths $\theta=0.1,0.5,$ and $0.9$. Vertical dotted lines indicate the first maxima of the GGM, with circles marking the corresponding QFI. We observe that the crossover from transient $\mathcal{F}_\theta(t)\sim t^4$ growth to approximately quadratic growth occurs as the multipartite entanglement approaches an oscillatory plateau. All the axes are dimensionless and are on a logarithmic scale.}
    \label{fig:ggm_qfi_comp}
\end{figure*}
Apart from providing the quantum benefit of QFI for different perturbations of the toric code Hamiltonian, we also present its correspondence with the generation of multipartite entanglement in the system, based on the generalized geometric measure (GGM) \cite{Bennett2000, Eisert2001, Verstraete2003, Wei2003, Ma2011, Hashemi2012} $\mathcal{G}(\ket{\psi(t)})$ of the time evolved state $\ket{\psi(t)}$ (see Appendix~\ref{app:ggm_definition}). In particular, for a given pure state $\ket{\psi}$ of $N$ sites, we compute the geometric multipartite entanglement \cite{Wei2003,Sen2010,ggm_shimony,Barnum_ggm_2} content as
\begin{align}
    \mathcal{G}_1(| \psi\rangle)=1-\max_{(A:B)\in\mathfrak{B}}[\xi_{A:B}^2 \mid \lVert A\rVert=1],
    \label{eq:G1_def}
\end{align}
where $\mathfrak{B}=\{(A:B) \mid A\cap B\!=\!\emptyset, \lVert A\cup B\rVert\!=\!N\}$ is the set of all bipartitions of the system, \(\xi_{A:B}\) is the largest Schmidt coefficient of the bipartition \(A:B\), and the maximization is performed only over single site bipartitions, i.e, with $\lVert A\rVert=1$. Note that a genuine measure of multipartite entanglement requires the maximization in Eq.~(\ref{eq:ggm_full}) over all bipartitions, $\lVert A \rVert \in \{1, \ldots, \lfloor N/2 \rfloor\}$, whose number grows exponentially with the system size. Hence, we restrict it to single-site density matrices, as $\Delta\mathcal{G} = \mathcal{G}_{1} - \mathcal{G} = 0$ due to the diagonal nature of the subsystem's density matrix \cite{LakkarajuToricDynamics24}, and use $\mathcal{G}_{1}$ and $\mathcal{G}$ interchangeably throughout. We show that for the optimal initial product state providing maximum temporal scaling of QFI in the transient regime, $\mathcal{F}_\theta(t)\sim t^4$ only until the multipartite entanglement is generated within the system. 

\subsection{Scaling of resources in quantum sensing}
\label{subsec:resource_scaling}

The metrological performance of a quantum many-body probe is evaluated by the scaling of the QFI with the resources such as the duration $t$ of the parameter encoding and the system size $N$ of the probe. For a probe comprising $N$ particles in a separable state with independent parameter encoding, the QFI is additive with at most linear growth, $\mathcal{F}_\theta \propto N$, whereas for an initially entangled probe, a quadratic growth, $\mathcal{F}_\theta \propto N^2$, termed the Heisenberg limit (HL), can be obtained~\cite{GLM04, GLM06}. Similarly, time plays an analogous role as a resource in the dynamical encoding protocols. While the independent encoding strategies give quadratic scaling with time $\mathcal{F}_\theta \sim t^2$, super-quadratic scaling, particularly up to $\mathcal{F}_\theta\sim t^4$, can be obtained in quantum systems at initial times, which reverts to the quadratic scaling at late times~\cite{PangBrun14}. 
It is thus natural to characterize the performance of a dynamical sensing protocol through the joint scaling ansatz,
\begin{equation}
    \mathcal{F}_\theta(N, t) \sim N^a\, t^b,
    \label{eq:joint-scaling}
\end{equation}
where $(a, b) = (1, 2)$ corresponds to the standard quantum limit (SQL)~\cite{BoixoPRL07, rams2018, Puig2025}. 
The temporal scaling of QFI with $b>2$ provides a quantum benefit and is generally obtained via a time-dependent Hamiltonian with a control Hamiltonian and feedback~\cite{PangJordan17}. We now demonstrate that a $t^4$ scaling of the QFI (quartic growth) can arise over a transient window, whenever the initial state can annihilate the prefactor of the quadratic term.  
A quartic scaling of the QFI in time is therefore a legitimate feature of {\it quantum} dynamics, without requiring additional resources or engineered time dependence.

In the non-equilibrium scenario, in which a probe state $\ket{\psi(0)}$ evolves unitarily under the time-independent parametrized Hamiltonian $H(\theta)$, so that $\ket{\psi_\theta(t)} = e^{-i H(\theta) t}\ket{\psi(0)}$. The QFI of the time evolved state, for $t \ll 1$, 
is given by~\cite{PangBrun14, PangJordan17}
\begin{align}
    &\mathcal{F}_\theta(t) = 4t^2\mathrm{Var}_{\ket{\psi_0}}(\partial_\theta H) + 4t^3\mathrm{Cov}_{\ket{\psi_0}}\!\left(\partial_\theta H,iC(\theta)\right) \nonumber \\ +& t^4\left[ \mathrm{Var}_{\ket{\psi_0}}\!\left(iC(\theta)\right) -\frac{4}{3}\mathrm{Cov}_{\ket{\psi_0}}\!\left( \partial_\theta H,[H,C(\theta)] \right) \right] + \mathcal{O}(t^5)
\end{align}
where $C(\theta)=[H(\theta),\partial_\theta H(\theta)]$ is the commutator,  $\mathrm{Var}_{\ket{\psi}}(A)=\langle\psi\lvert A^2\rvert\psi\rangle - \langle\psi\lvert A\rvert\psi\rangle^2$ is the variance of a Hermitian operator $A$ and $\mathrm{Cov}_{\ket{\psi}}(A,B)=\frac{1}{2}\bra{\psi}AB+BA\ket{\psi}-\bra{\psi}A\ket{\psi}\bra{\psi}B\ket{\psi}$ is the symmetrized covariance between Hermitian operators $A$ and $B$.
It immediately follows that a super-quadratic transient scaling can arise only if the leading term vanishes, i.e., if
\begin{equation}
    \mathrm{Var}_{\ket{\psi_0}}(\partial_\theta H(\theta)) = 0,
    \label{eq:t4-condition}
\end{equation}
which is fulfilled precisely when the initial state is an eigenstate of $\partial_\theta H(\theta)$. Interestingly, whenever Eq.~(\ref{eq:t4-condition}) holds, the $\mathcal{O}(t^3)$ contribution to the QFI vanishes identically as well, so that the first surviving correction is of fourth order,
\begin{equation}
    F_Q(t) = t^4\, \mathrm{Var}_{\ket{\psi_0}}\!\left( i\left[ H(\theta), \partial_\theta H(\theta) \right] \right) + \mathcal{O}(t^5).
    \label{eq:t4-qfi}
\end{equation}
Therefore, Eq.~(\ref{eq:t4-condition}) provides a simple recipe for the state preparation to obtain quartic scaling of QFI in the dynamical encoding. An eigenstate of $\partial_\theta H_\theta$ with a large fluctuation of the commutator $i[H_\theta, \partial_\theta H_\theta]$ results in the QFI starting off on a $t^4$ trajectory initially. \\

\textit{Observation 1. Optimal initial product states.} Each perturbation commutes with its own differential, i.e., $[H_X(\theta),\partial_\theta H_X(\theta)]=0$ for all the models $X\equiv\{\text{CC, Ising, Mag}\}$, so that every eigenstate of $H_X(\theta)$ is also an eigenstate of $\partial_\theta H_X(\theta)$ and hence satisfies Eq.~(\ref{eq:t4-condition}). The product probes $\ket{\psi(0)}=\ket{\uparrow}^{\otimes N}$ (or $\ket{\downarrow}^{\otimes N}$) for $X\in\{\text{CC},\text{Mag}\}$, and $\ket{\psi(0)}=\ket{\pm}^{\otimes N}$ for $X\equiv\text{Ising}$, therefore exhibit the quartic growth of Eq.~(\ref{eq:t4-qfi}), as shown in Fig.~\ref{fig:ggm_qfi_comp}. We further find the increase to be linear in the system size, $\mathcal{F}_\theta(t)\sim N t^4$ for $N=8, 12, 16$ and $20$, i.e., $(a,b)=(1,4)$ in Eq.~(\ref{eq:joint-scaling}). These are precisely the probes employed in the disordered case of Sec.~\ref{sec:disorderQFI}, so that a single family of separable states serves both parts of the study.

\begin{figure}
    \centering
    \includegraphics[width=\columnwidth]{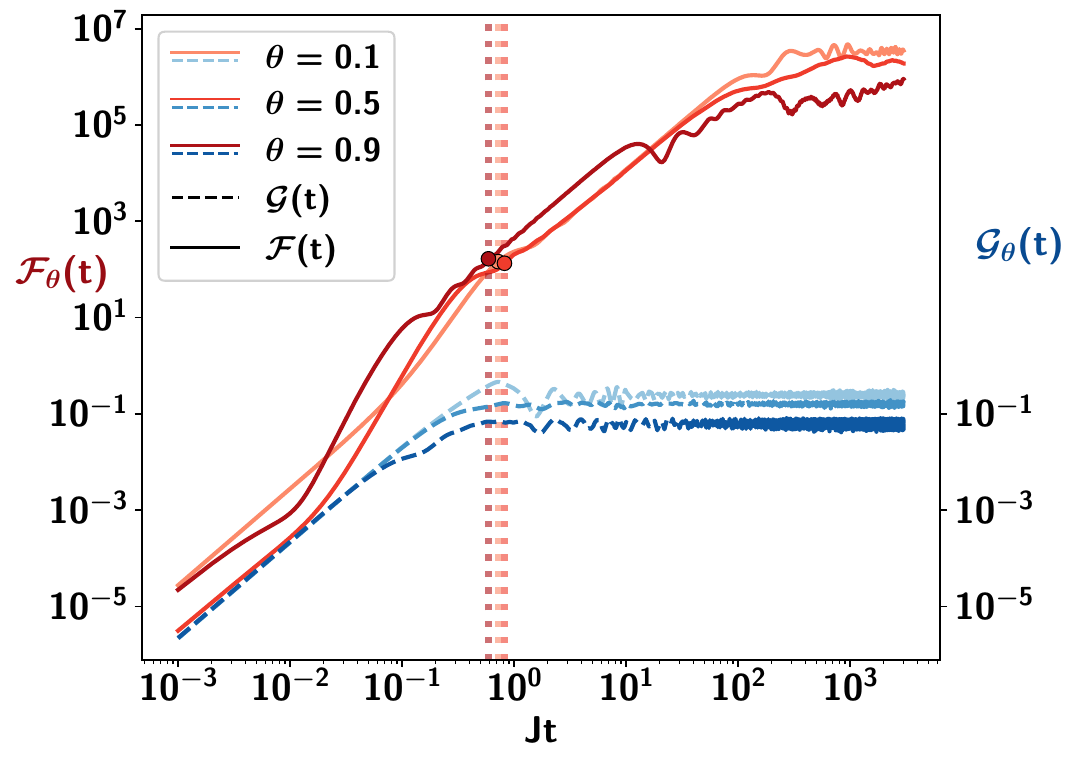 }
    \caption{{\bf 
    Trends of QFI and GGM for a non-optimal initial product probe.} The QFI $\mathcal{F}_\theta(t)$ (left ordinate, solid red curves) and GGM $\mathcal{G}_\theta(t)$, dashed blue curves), against time $t$ (abscissa) for the Castelnovo-Chamon perturbation with $N=20$ and site-independent interaction $J$. The initial state is $\ket{\phi}^{\otimes N}$, with $\ket{\phi}=\cos(\pi/12)\ket{\uparrow}+e^{i 25\pi/18}\sin(\pi/12)\ket{\downarrow}$, as an arbitrarily chosen non-optimal state. Different shades represent different perturbation strengths $\theta=0.1,0.5,$ and $0.9$. The QFI initially grows quadratically, develops a transient window of quartic growth, which returns to quadratic growth when the multipartite entanglement saturates with small oscillations. All the axes are dimensionless and on a logarithmic scale.}
    \label{fig:qfi_loglog_30_250}
\end{figure}

As expected, at long times, the QFI of a state evolving under a time-independent Hamiltonian is quadratic in $t$ together with bounded oscillatory contributions~\cite{PangBrun14}, so that the asymptotic growth inevitably relaxes back to the $t^2$ envelope. The operationally relevant quantities are, therefore, both the transient exponent $b > 2$ in Eq.~\eqref{eq:joint-scaling} and the temporal window $t\lesssim 1$ over which it persists.

\textit{Remark}. The Castelnovo-Chamon perturbation shows intricate behavior for non-optimal initial states. Specifically for $\ket{\psi(0)} =  \ket{\phi}^{\otimes N}$ with $|\phi\rangle = \cos(\omega/2)\ket{\uparrow}+e^{i\kappa}\sin(\omega/2)\ket{\downarrow}$ (with $\omega\neq0,\pi$), the corresponding QFI $\mathcal{F}_\theta(t)$ exhibits a qualitatively different behavior. Specifically, in an initial transient window, $ \mathcal{F}_\theta (t)\sim t^2$ scaling, followed by $t^4$ scaling till $t\lesssim 1$, before relaxing back to $\sim t^2$ growth and eventually saturating into the oscillatory regime. 

These states are not eigenstates of $\partial_\theta H(\theta)$ at $t=0$ and so do not exhibit $t^4$ scaling from the start, but the dynamics drives them close to the relevant eigenspace of $\partial_\theta H(\theta)$ with suppressed variance $\mathrm{Var}_{\ket{\psi_0}}(\partial_\theta H(\theta))$, and hence a $t^4$-scaling is observed in the transient time (see Fig.~\ref{fig:qfi_loglog_30_250} for $\omega = \frac{\pi}{6},\kappa = \frac{25\pi}{18}$ chosen arbitrarily). Note that even when the disorder is introduced in this model, the same $t^4$ scaling in the quenched average QFI can be obtained, thereby exhibiting the system's robustness under preparation imperfections.

The standard connection between QFI and multipartite entanglement (ME) concerns the sensitivity of a given $k$-separable probe to a phase encoded by local rotations~\cite{HyllusPRA12, TothPRA12}. For an $N$-qubit $k$-separable probe $\rho$ undergoing $\rho_\theta=e^{-i\theta G}\rho e^{i\theta G}$, the QFI satisfies $\mathcal{F}_\theta[\rho_\theta]\leq4\mathrm{Var}_{\rho}(G)$, with equality for pure state probes, where $G$ is the local normalised generator of the encoding. For $G=\frac{1}{2}\sum_{j=1}^{N}\boldsymbol{n}_j\cdot\boldsymbol{\sigma}_j$, with $|\boldsymbol{n}_j|=1$, fully separable probes with $k=N$ obey $\mathcal{F}_\theta\leq N$, while for multipartite entangled states, larger system-size scaling for QFI can be obtained. In this setting, bounds on the entanglement already present in the input probe constrain its attainable sensitivity for the chosen generator, while the local encoding itself leaves that entanglement unchanged~\cite{HyllusPRA12, TothPRA12}.
\begin{figure*}
    \centering
    \includegraphics[width=1\linewidth]{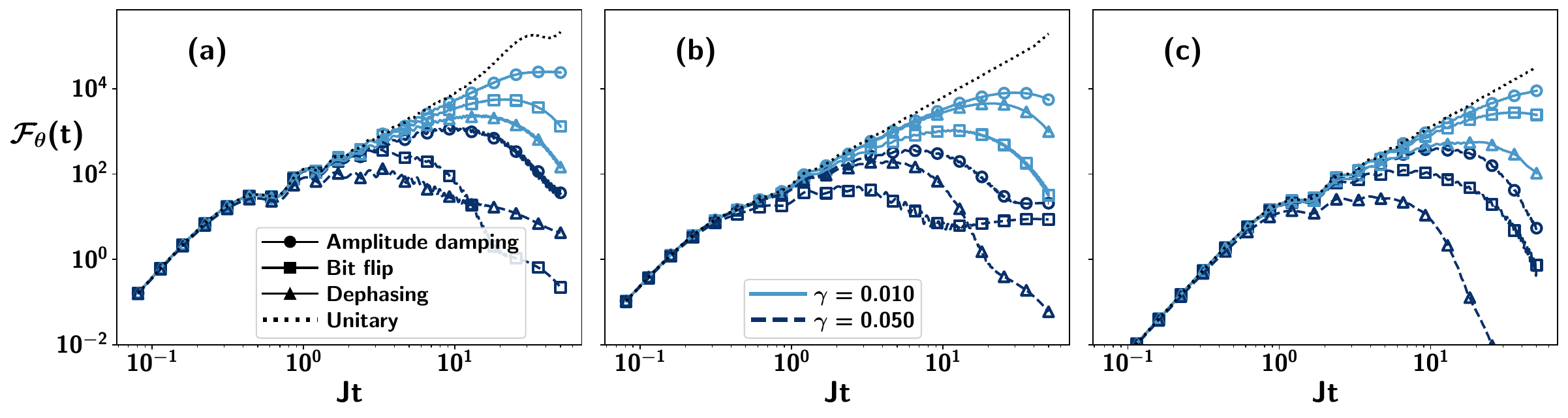}
    \caption{{\bf Persistence of transient quartic QFI growth under local noise.} The QFI (ordinate) against the time $t$ (abscissa) for (a) Castelnovo-Chamon, (b) Ising, and (c) magnetic field perturbations on horizontal edges of the toric code without disorder, with $N=8$, and perturbation strength $\theta=0.5$. The initial probes are $\ket{\uparrow}^{\otimes N}$ for (a) and (c), and $\ket{+}^{\otimes N}$ for (b). 
    Circles, squares, and triangles represent independent local amplitude damping, bit-flip, and dephasing noise, respectively, while a dotted curve corresponds to the unitary evolution  (noiseless scenario).  We choose two noise strengths $\gamma=0.01$ (light line) and $0.05$ (dark dashed line). 
    The noisy and noiseless curves nearly coincide at short times, retaining $\mathcal{F}_\theta(t)\sim t^4$ scaling. At later times, noise suppresses the QFI, with stronger noise affecting earlier times, decreasing the QFI and a greater reduction in sensitivity. For a given noise strength, the amplitude damping channel has the least effect for all the perturbations, while dephasing has the largest effect for (a) and (c) and the bit-flip for (b). 
    All the axes are dimensionless and on a logarithmic scale.}
    \label{fig:noise_qfi_comp}
\end{figure*}
The relationship between ME and QFI investigated here is different. Our parameter $\theta$ is a coupling in the interacting Hamiltonian $H(\theta)$, and the evolution $U_\theta(t)=e^{-iH_{TC}(\theta)t}$ simultaneously encodes the parameter and generates entanglement from an initially separable probe. The QFI is then
\begin{equation}
    \mathcal{F}_\theta(t)=4\mathrm{Var}_{\ket{\psi_0}}\!\left(g_\theta(t)\right),
    \qquad g_\theta(t)=iU_\theta^\dagger(t)\partial_\theta U_\theta(t),
    \label{eq:qfi_dynamical_generator}
\end{equation}
where the effective generator depends on the encoding time, the interacting dynamics, and is non-local. We, therefore,  examine how the temporal growth of the QFI is related to the created multipartite entanglement in $\ket{\psi_\theta(t)}$, rather than using QFI under local rotations to witness the entanglement of a prepared state. In particular, for the initial product states identified above, we investigate whether the duration of the transient quartic regime is associated with the interval over which entanglement grows. We find that this is indeed true. \\

\textit{Observation 2. Connecting the QFI with the multipartite entanglement}.  
To quantify this dynamical correspondence, we compare the time dependence of the $\mathcal{F}_\theta(t)$ with that of the multipartite entanglement $\mathcal{G}_\theta(t)$ produced from the initial product probes. Notice that we drop the subscript $1$ due to the diagonal nature of subsystem density matrices \cite{LakkarajuToricDynamics24}.

We compute the $\mathcal{G}_\theta(t)$ for each optimal initial state as it evolves. We observe that it shows a similar profile to the $\mathcal{F}_\theta(t)$, with an initial region with close to $t^2$ scaling, followed by saturation and oscillation at long times. Figure~\ref{fig:ggm_qfi_comp} overlays the two quantities on a log-log scale for all three models and several values of $\theta$, marking the $\mathcal{F}_\theta(t)$ at the time of the GGM's first minimum. We further observe that the QFI (solid) and GGM (dashed) rise together during the initial transient period; the $\mathcal{G}_\theta(t)$ then moves into an oscillatory plateau while the QFI continues climbing along a $t^2$ trajectory. The time at which the $\mathcal{G}_\theta(t)$ plateau starts consistently coincides with the QFI's $t^4$ window changing into the $t^2$ window. It is clearest in the $\mathrm{CC}$ and $\mathrm{Ising}$ cases, where the QFI curves for different $\theta$ visibly change near the corresponding point (marked by circles in Fig. \ref{fig:ggm_qfi_comp}), while its change is not that sharp for the $\mathrm{Mag}$ case. In other words, the QFI's $t^4$-scaling window persists for as long as the entanglement in the state is still building up, and ends once entanglement growth saturates. It indicates a direct link between the transient window of metrological advantage and the increase of the entanglement generated by the encoding dynamics.

\section{Robustness of dynamical sensing under local noise}
\label{sec:noise}

We have already established that in certain parameter regimes and time domains, imperfections during the preparation of the Kitaev toric code model can enhance metrological precision in terms of the QFI. Let us now investigate how the QFI is affected when the system interacts with an environment during dynamics~\cite{Wasilewski_2010, Escher2011, DemkowiczDobrzaski2012, Koodyski2013, Alipour_2014, haase2016_open_review, Falaye2017}. In other words, let us examine whether the transient quartic growth discussed in Sec.~\ref{sec:dynamics} during the coherent evolution of an initially separable probe persists when the probe interacts with an environment. In order to address this question, we consider local dephasing, bit-flip, and amplitude damping noise \cite{Preskill1998, lidar_2020_lecture} that act locally at each site during parameter encoding. This implies that noise at each site acts independently, i.e., noise is uncorrelated. Due to the bath ($B$) described by the Lindblad operators $\{L_k^{}\}$ with noise strength $\gamma$ on each site, the dynamics of the state can be described by the Gorini-Kossakowski-Sudarshan-Lindblad (GKSL) master equation~\cite{open_quan_book, Rivas2012, lidar_2020_lecture} as 
\begin{align}
    \frac{d\rho_\theta}{dt}
    = -i[H_{TC}(\theta),\rho_\theta] + J\gamma \sum_{k=1}^{N}\mathcal{D}[L_k] (\rho_\theta), \label{eq:gksl}
\end{align}
with $\mathcal{D}[L](\rho) = L\rho L^\dagger-\frac{1}{2}\{L^\dagger L,\rho\}$. By taking the initial state  as described before, the dynamical state is obtained for a fixed noise model specified by Lindblad operators, which can then be used to compute the QFI as a function of time and noise parameters.

We consider the effects of three types of noise, namely
\begin{enumerate}
    \item [a.]  {\it Dephasing noise}. Local dephasing in the $z$-direction is described by $L_k=\sigma_k^z$. In the absence of Hamiltonian evolution, the dissipator leaves the populations in the $\sigma^z$ eigenbasis unchanged and becomes responsible for the decay of the off-diagonal matrix elements as $e^{-2\gamma t}$. Although the initial $z$-polarized probes are unaffected by dephasing alone, the toric code interactions generate coherences on which the noise acts subsequently.
    \item [b.] {\it Bit-flip noise.} This flips the basis $\{\ket{\uparrow}_k, \ket{\downarrow}_k\}$, and is generated by $L_k=\sigma_k^x$, and the corresponding dissipator is $\mathcal{D}[\sigma_k^x](\rho) = \gamma(\sigma_k^x\rho\sigma_k^x-\rho)$, describing symmetric transitions between $\ket{\uparrow}_k$ and $\ket{\downarrow}_k$, and keeps the $\ket{+}_j$ state invariant. Both the dephasing and bit-flip channels are unital noises, as they leave the identity operator invariant.
    \item [c.] {\it Amplitude damping noise.} In this non-unital noise, $L_k=\sigma_k^- = \ket{\downarrow}_k\bra{\uparrow}_k$, which transfers population from $\ket{\uparrow}_k$ to $\ket{\downarrow}_k$ at rate $\gamma$ and leaves $\ket{\downarrow}_k$ unaffected for each $k$ in the absence of the Hamiltonian. Unlike the two Pauli channels, it selects a preferred local state for relaxation in the physical spin basis.
\end{enumerate}

The GKLS master equation is solved numerically, in general, with a fourth-order Runge-Kutta (RK4) scheme. Since the evolved probe is mixed, we restrict to small system-sizes and the corresponding QFI is evaluated using the mixed-state expression in Eq.~\eqref{eq:mixed_QFI_expr}, with the sum restricted to eigenvalue pairs satisfying $\lambda_j+\lambda_k>0$~\cite{LiuCTP14, LiuJPA20}, which is set to $\lambda_j+\lambda_k>\epsilon$, where $\epsilon$ is the numerical threshold of error in the RK4 method.

{\it Persistence of the transient quartic regime.}
The temporal scaling of QFI $\mathcal{F}_\theta(t)\sim t^4$ is observed to be robust to all the local noise models considered here for the transient time. The QFI curves nearly coincide in the initial interval with $t\ll1$, specifically when $10^{-3} \lesssim t \lesssim 10^{-1} $ for $\gamma = 0.01, 0.05$, irrespective of the perturbations, as depicted in Fig. \ref{fig:noise_qfi_comp}. The entire investigations for both unital and non-unital channels indicate that the transient scaling of QFI remains unaffected with noise. Thus topological systems turn out to be robust, both against the imperfections of the lattice sites and the noisy environment, especially in the transient regimes. However, as the system evolves further, the QFI gets affected due to noise although the metrological benefit in terms of the scaling with time is typically obtained in the transient domain. Let us discuss briefly how the QFI behaves beyond small times.

{\it Behaviour beyond the transient regime.}
As the encoding time increases beyond the initial quartic window, the curves separate and develop a pronounced dependence on the noise rate. Increasing $\gamma$ from $0.01$ to $0.05$ suppresses the later-time QFI and generally brings its turnover to earlier times. The onset and extent of this suppression depend on the Hamiltonian and the channel. Interestingly, the effect of amplitude damping is less than that of the dephasing and the bit-flip noises for the three perturbations, suggesting higher robustness of the QFI for the non-unital noise. Specifically, for $N=8$, $\mathcal{F}_\theta(t)$ decreases with time for $t\geq30$ for $\gamma=0.01$ and $t\geq10$ for $\gamma=0.05$ for all models, $\mathrm{CC},\mathrm{Ising} \text{ and, } \mathrm{Mag}$. Interestingly, the dephasing deteriorates the QFI more than the bit-flip only for the $\mathrm{CC}$ and the $\mathrm{Mag}$ perturbations, whereas the bit-flip decreases the QFI faster than the dephasing noise in the $\mathrm{Ising}$ perturbations. Therefore, the effect of noise on QFI at later times is greater for the unital channels, which is faster when the noise directly affects the encoding perturbation.

The useful sensing interval must therefore be assessed from the magnitude of the QFI as well as its temporal exponent. The observed quartic regime survives the weak local noise considered here, while longer encoding times expose substantial channel-dependent losses. This short-time persistence concerns dynamically generated sensitivity in a product-state protocol, and does not establish topological protection of the metrological precision.

\section{Conclusion}
\label{sec:conclusion}

Topologically ordered systems are shown to be resilient against sufficiently weak perturbations in equilibrium, although this robustness need not extend to their nonequilibrium dynamics. In this work, we investigated the dynamical estimation of the strengths of three perturbations of the Kitaev toric code, the Castelnovo-Chamon deformation, nearest-neighbor Ising interactions, and a magnetic field along horizontal edges, using suitably chosen initially separable probes. This framework allowed us to systematically examine the roles of the toric-code interactions, spatial disorder, and dynamically generated multipartite entanglement in the encoding process, and also the impact of local environmental noise.

We demonstrated that disorder in the toric-code couplings can enhance the quenched average quantum Fisher information (QFI), a performance quantifier of quantum sensing, in appropriate parameter regimes. For the initial probes considered, only randomness sampled from a Gaussian distribution in the star couplings contributes to the QFI for the Castelnovo--Chamon and horizontal-field perturbations, while only randomly chosen plaquette couplings affect the QFI for the Ising perturbation. For the former two perturbations, weak perturbations exhibit an enhancement at early times that diminishes at later times, although sufficiently strong perturbations can exhibit enhancement at both early and late encoding times for the Ising case. 
These results demonstrate that disorder can have a constructive impact in the dynamical sensing of finite toric-code systems, which we referred to as {\it disorder-induced QFI}.

In the absence of disorder, we further showed that the QFI can exhibit transient quartic growth for suitably chosen initial product probes satisfying specific conditions for each of the three perturbations. We found that the crossover from the quartic regime to an approximately quadratic temporal scaling happens when the multipartite entanglement generated during the encoding dynamics approaches an oscillatory plateau. This correspondence connects the duration of the transient quartic scaling with the generated entanglement growth.

To determine the impact of environmental interactions during the encoding dynamics, we considered local dephasing, bit-flip, and amplitude-damping noise and obtained the dynamical state by solving the Gorini-Kossakowski-Sudarshan-Lindblad master equation. We found that the systems retain their transient quartic scaling  in the presence of all three noisy channels, demonstrating the robustness of this metrological regime against decoherence. At longer times, however, the QFI decays, with the decay behavior depending on the noisy channel parameters. Among the channels considered, amplitude damping exhibits the greatest robustness of the QFI.

Our results demonstrate that the choice of the initial probe and the non-commuting interactions of the toric code provide key ingredients for controlling the temporal response of a quantum sensor, while spatial disorder can serve as an additional resource for enhancing sensitivity in suitable regimes. It would be interesting to determine the extent to which this disorder-induced enhancement persists when the disorder realization is unknown and also when noisy encoding dynamics is incorporated.

\acknowledgments
We  acknowledge  the cluster computing facility at Harish-Chandra Research Institute and  support from the project entitled ``Technology Vertical - Quantum Communication'' under the National Quantum Mission of the Department of Science and Technology (DST)  (Sanction Order No. DST/QTC/NQM/QComm/$2024/2$ (G)).
This project has been funded by the Caritro Foundation. This work was supported by the Provincia Autonoma di Trento, and Q@TN, the joint lab between the University of Trento, FBK—Fondazione Bruno Kessler, INFN—National Institute for Nuclear Physics, and CNR—National Research Council, Italy.

\bibliography{topo_q}

@article{giovannetti_nature,
  title = {Advances in quantum metrology},
  author = {Giovannetti, Vittorio and Lloyd, Seth and Maccone, Lorenzo},
  journal = {Nature Photonics},
  volume = {5},
  issue = {4},
  pages = {222-229},
  numpages = {7},
  year = {2011},
  month = {April},
  publisher = {},
  doi = {10.1038/nphoton.2011.35},
  url = {https://doi.org/10.1038/nphoton.2011.35}
}

@article{utkarsh2021,
  title = {Driving Enhanced Quantum Sensing in Partially Accessible Many-Body Systems},
  author = {Mishra, Utkarsh and Bayat, Abolfazl},
  journal = {Phys. Rev. Lett.},
  volume = {127},
  issue = {8},
  pages = {080504},
  numpages = {7},
  year = {2021},
  month = {Aug},
  publisher = {American Physical Society},
  doi = {10.1103/PhysRevLett.127.080504},
  url = {https://link.aps.org/doi/10.1103/PhysRevLett.127.080504}
}

@article{rams2018,
  title = {At the Limits of Criticality-Based Quantum Metrology: Apparent Super-Heisenberg Scaling Revisited},
  author = {Rams, Marek M. and Sierant, Piotr and Dutta, Omyoti and Horodecki, Pawe\l{} and Zakrzewski, Jakub},
  journal = {Phys. Rev. X},
  volume = {8},
  issue = {2},
  pages = {021022},
  numpages = {16},
  year = {2018},
  month = {Apr},
  publisher = {American Physical Society},
  doi = {10.1103/PhysRevX.8.021022},
  url = {https://link.aps.org/doi/10.1103/PhysRevX.8.021022}
}

@article{Sensing_RMP_2017,
  title = {Quantum sensing},
  author = {Degen, C. L. and Reinhard, F. and Cappellaro, P.},
  journal = {Rev. Mod. Phys.},
  volume = {89},
  issue = {3},
  pages = {035002},
  numpages = {39},
  year = {2017},
  month = {Jul},
  publisher = {American Physical Society},
  doi = {10.1103/RevModPhys.89.035002},
  url = {https://link.aps.org/doi/10.1103/RevModPhys.89.035002}
}

@article{BoixoPRL07,
  title = {Generalized Limits for Single-Parameter Quantum Estimation},
  author = {Boixo, Sergio and Flammia, Steven T. and Caves, Carlton M. and Geremia, JM},
  journal = {Phys. Rev. Lett.},
  volume = {98},
  issue = {9},
  pages = {090401},
  numpages = {4},
  year = {2007},
  month = {Feb},
  publisher = {American Physical Society},
  doi = {10.1103/PhysRevLett.98.090401},
  url = {https://link.aps.org/doi/10.1103/PhysRevLett.98.090401}
}

@article{Puig2025,
  title = {From Dynamical to Steady-State Many-Body Metrology: Precision Limits and Their Attainability with Two-Body Interactions},
  author = {Puig, Ricard and Sekatski, Pavel and Erdman, Paolo Andrea and Abiuso, Paolo and Calsamiglia, John and Perarnau-Llobet, Mart\'{\i}},
  journal = {PRX Quantum},
  volume = {6},
  issue = {3},
  pages = {030309},
  numpages = {35},
  year = {2025},
  month = {Jul},
  publisher = {American Physical Society},
  doi = {10.1103/PRXQuantum.6.030309},
  url = {https://link.aps.org/doi/10.1103/PRXQuantum.6.030309}
}

@article{RahmaniChamon10,
  title = {Exact results on the quench dynamics of the entanglement entropy in the toric code},
  author = {Rahmani, Armin and Chamon, Claudio},
  journal = {Phys. Rev. B},
  volume = {82},
  issue = {13},
  pages = {134303},
  numpages = {14},
  year = {2010},
  month = {Oct},
  publisher = {American Physical Society},
  doi = {10.1103/PhysRevB.82.134303},
  url = {https://link.aps.org/doi/10.1103/PhysRevB.82.134303}
}

@article{Paris09,
  author  = {Paris, Matteo G. A.},
  title   = {Quantum Estimation for Quantum Technology},
  journal = {Int. J. Quantum Inf.},
  volume  = {7},
  number  = {Supplement},
  pages   = {125--137},
  year    = {2009},
  doi     = {10.1142/S0219749909004839}
}

@article{LiuCTP14,
  author  = {Liu, Jing and Jing, Xiao-Xing and Zhong, Wei and Wang, Xiao-Guang},
  title   = {Quantum Fisher Information for Density Matrices with Arbitrary Ranks},
  journal = {Commun. Theor. Phys.},
  volume  = {61},
  number  = {1},
  pages   = {45--50},
  year    = {2014},
  doi     = {10.1088/0253-6102/61/1/08}
}

@article{LiuJPA20,
  author  = {Liu, Jing and Yuan, Haidong and Lu, Xiao-Ming and Wang, Xiaoguang},
  title   = {Quantum Fisher Information Matrix and Multiparameter Estimation},
  journal = {J. Phys. A},
  volume  = {53},
  number  = {2},
  pages   = {023001},
  year    = {2020},
  doi     = {10.1088/1751-8121/ab5d4d}
}

@article{GLM04,
  author  = {Giovannetti, Vittorio and Lloyd, Seth and Maccone, Lorenzo},
  title   = {Quantum-Enhanced Measurements: Beating the Standard Quantum Limit},
  journal = {Science},
  volume  = {306},
  number  = {5700},
  pages   = {1330--1336},
  year    = {2004},
  doi     = {10.1126/science.1104149}
}

@article{GLM06,
  author  = {Giovannetti, Vittorio and Lloyd, Seth and Maccone, Lorenzo},
  title   = {Quantum Metrology},
  journal = {Phys. Rev. Lett.},
  volume  = {96},
  pages   = {010401},
  year    = {2006},
  doi     = {10.1103/PhysRevLett.96.010401}
}

@article{PangBrun14,
  author  = {Pang, Shengshi and Brun, Todd A.},
  title   = {Quantum Metrology for a General Hamiltonian Parameter},
  journal = {Phys. Rev. A},
  volume  = {90},
  pages   = {022117},
  year    = {2014},
  doi     = {10.1103/PhysRevA.90.022117}
}

@article{PangJordan17,
  author  = {Pang, Shengshi and Jordan, Andrew N.},
  title   = {Optimal Adaptive Control for Quantum Metrology with Time-Dependent Hamiltonians},
  journal = {Nat. Commun.},
  volume  = {8},
  pages   = {14695},
  year    = {2017},
  doi     = {10.1038/ncomms14695}
}

@article{HyllusPRA12,
  author  = {Hyllus, Philipp and Laskowski, Wieslaw and Krischek, Roland and Schwemmer, Christian and Wieczorek, Witlef and Weinfurter, Harald and Pezz{\'e}, Luca and Smerzi, Augusto},
  title   = {Fisher Information and Multiparticle Entanglement},
  journal = {Phys. Rev. A},
  volume  = {85},
  pages   = {022321},
  year    = {2012},
  doi     = {10.1103/PhysRevA.85.022321}
}

@article{TothPRA12,
  author  = {T{\'o}th, G{\'e}za},
  title   = {Multipartite Entanglement and High-Precision Metrology},
  journal = {Phys. Rev. A},
  volume  = {85},
  pages   = {022322},
  year    = {2012},
  doi     = {10.1103/PhysRevA.85.022322}
}

@article{Kitaev03,
  author  = {Kitaev, A. Yu.},
  title   = {Fault-Tolerant Quantum Computation by Anyons},
  journal = {Ann. Phys.},
  volume  = {303},
  number  = {1},
  pages   = {2--30},
  year    = {2003},
  doi     = {10.1016/S0003-4916(02)00018-0}
}

@article{CastelnovoChamon08,
  title = {Quantum topological phase transition at the microscopic level},
  author = {Castelnovo, Claudio and Chamon, Claudio},
  journal = {Phys. Rev. B},
  volume = {77},
  issue = {5},
  pages = {054433},
  numpages = {14},
  year = {2008},
  month = {Feb},
  publisher = {American Physical Society},
  doi = {10.1103/PhysRevB.77.054433},
  url = {https://link.aps.org/doi/10.1103/PhysRevB.77.054433}
}

@article{santra2014,
  title = {Local convertibility of the ground state of the perturbed toric code},
  author = {Santra, Siddhartha and Hamma, Alioscia and Cincio, Lukasz and Subasi, Yigit and Zanardi, Paolo and Amico, Luigi},
  journal = {Phys. Rev. B},
  volume = {90},
  issue = {24},
  pages = {245128},
  numpages = {18},
  year = {2014},
  month = {Dec},
  publisher = {American Physical Society},
  doi = {10.1103/PhysRevB.90.245128},
  url = {https://link.aps.org/doi/10.1103/PhysRevB.90.245128}
}

@article{braunstein1994,
  title = {Statistical distance and the geometry of quantum states},
  author = {Braunstein, Samuel L. and Caves, Carlton M.},
  journal = {Phys. Rev. Lett.},
  volume = {72},
  issue = {22},
  pages = {3439--3443},
  numpages = {0},
  year = {1994},
  publisher = {American Physical Society},
  doi = {10.1103/PhysRevLett.72.3439},
  url = {https://link.aps.org/doi/10.1103/PhysRevLett.72.3439},
  month = {May}
}

@book{Helstrom1976,
  author    = {Helstrom, Carl W.},
  title     = {Quantum Detection and Estimation Theory},
  publisher = {Academic Press},
  address   = {New York},
  year      = {1976},
  series    = {Mathematics in Science and Engineering},
  volume    = {123},
  isbn = {0-12-340050-3}
}

@book{Holevo1982,
  author    = {Holevo, Alexander S.},
  title     = {Probabilistic and Statistical Aspects of Quantum Theory},
  publisher = {North-Holland Publishing Company},
  address   = {Amsterdam},
  year      = {1982},
  series    = {North-Holland Series in Statistics and Probability},
  volume    = {1},
  isbn = {0-444-86333-8}
}

@article{Wei2003,
  title = {Geometric measure of entanglement and applications to bipartite and multipartite quantum states},
  author = {Wei, Tzu-Chieh and Goldbart, Paul M.},
  journal = {Phys. Rev. A},
  volume = {68},
  issue = {4},
  pages = {042307},
  numpages = {12},
  year = {2003},
  month = {Oct},
  publisher = {American Physical Society},
  doi = {10.1103/PhysRevA.68.042307},
  url = {https://link.aps.org/doi/10.1103/PhysRevA.68.042307}
}

@article{Sen2010,
  title = {Channel capacities versus entanglement measures in multiparty quantum states},
  author = {Sen(De), Aditi and Sen, Ujjwal},
  journal = {Phys. Rev. A},
  volume = {81},
  issue = {1},
  pages = {012308},
  numpages = {6},
  year = {2010},
  month = {Jan},
  publisher = {American Physical Society},
  doi = {10.1103/PhysRevA.81.012308},
  url = {https://link.aps.org/doi/10.1103/PhysRevA.81.012308}
}

@article{ggm_shimony,
author = {Shimony, Abner},
title = {Degree of Entanglement},
journal = {Annals of the New York Academy of Sciences},
volume = {755},
number = {1},
pages = {675-679},
doi = {https://doi.org/10.1111/j.1749-6632.1995.tb39008.x},
url = {https://nyaspubs.onlinelibrary.wiley.com/doi/abs/10.1111/j.1749-6632.1995.tb39008.x},
year = {1995}
}

@article{Barnum_ggm_2,
doi = {10.1088/0305-4470/34/35/305},
url = {https://dx.doi.org/10.1088/0305-4470/34/35/305},
year = {2001},
month = {aug},
publisher = {},
volume = {34},
number = {35},
pages = {6787},
author = {H Barnum and N Linden},
title = {Monotones and invariants for multi-particle quantum
states},
journal = {Journal of Physics A: Mathematical and General}
}

@article{Bravyi2010,
  title = {Topological quantum order: Stability under local perturbations},
  volume = {51},
  ISSN = {1089-7658},
  url = {http://dx.doi.org/10.1063/1.3490195},
  DOI = {10.1063/1.3490195},
  number = {9},
  journal = {Journal of Mathematical Physics},
  publisher = {AIP Publishing},
  author = {Bravyi,  Sergey and Hastings,  Matthew B. and Michalakis,  Spyridon},
  year = {2010},
  month = {Sep}
}

@article{Konar_2022,
  title = {Designing robust quantum refrigerators in disordered spin models},
  author = {Konar, Tanoy Kanti and Ghosh, Srijon and Pal, Amit Kumar and Sen(De), Aditi},
  journal = {Phys. Rev. A},
  volume = {105},
  issue = {2},
  pages = {022214},
  numpages = {12},
  year = {2022},
  month = {Feb},
  publisher = {American Physical Society},
  doi = {10.1103/PhysRevA.105.022214},
  url = {https://link.aps.org/doi/10.1103/PhysRevA.105.022214}
}

@article{Ghosh_2020,
  title = {Enhancement in the performance of a quantum battery by ordered and disordered interactions},
  author = {Ghosh, Srijon and Chanda, Titas and Sen(De), Aditi},
  journal = {Phys. Rev. A},
  volume = {101},
  issue = {3},
  pages = {032115},
  numpages = {11},
  year = {2020},
  month = {Mar},
  publisher = {American Physical Society},
  doi = {10.1103/PhysRevA.101.032115},
  url = {https://link.aps.org/doi/10.1103/PhysRevA.101.032115}
}

@article{Fort2005,
  title = {Effect of Optical Disorder and Single Defects on the Expansion of a Bose-Einstein Condensate in a One-Dimensional Waveguide},
  author = {Fort, C. and Fallani, L. and Guarrera, V. and Lye, J. E. and Modugno, M. and Wiersma, D. S. and Inguscio, M.},
  journal = {Phys. Rev. Lett.},
  volume = {95},
  issue = {17},
  pages = {170410},
  numpages = {4},
  year = {2005},
  month = {Oct},
  publisher = {American Physical Society},
  doi = {10.1103/PhysRevLett.95.170410},
  url = {https://link.aps.org/doi/10.1103/PhysRevLett.95.170410}
}

@article{Roati2008,
  title = {Anderson localization of a non-interacting Bose–Einstein condensate},
  volume = {453},
  ISSN = {1476-4687},
  url = {http://dx.doi.org/10.1038/nature07071},
  DOI = {10.1038/nature07071},
  number = {7197},
  journal = {Nature},
  publisher = {Springer Science and Business Media LLC},
  author = {Roati,  Giacomo and D’Errico,  Chiara and Fallani,  Leonardo and Fattori,  Marco and Fort,  Chiara and Zaccanti,  Matteo and Modugno,  Giovanni and Modugno,  Michele and Inguscio,  Massimo},
  year = {2008},
  month = {June},
  pages = {895–898}
}

@article{Fallani_2007,
  title = {Ultracold Atoms in a Disordered Crystal of Light: Towards a Bose Glass},
  author = {Fallani, L. and Lye, J. E. and Guarrera, V. and Fort, C. and Inguscio, M.},
  journal = {Phys. Rev. Lett.},
  volume = {98},
  issue = {13},
  pages = {130404},
  numpages = {4},
  year = {2007},
  month = {Mar},
  publisher = {American Physical Society},
  doi = {10.1103/PhysRevLett.98.130404},
  url = {https://link.aps.org/doi/10.1103/PhysRevLett.98.130404}
}

@article{White_2009,
  title = {Strongly Interacting Bosons in a Disordered Optical Lattice},
  author = {White, M. and Pasienski, M. and McKay, D. and Zhou, S. Q. and Ceperley, D. and DeMarco, B.},
  journal = {Phys. Rev. Lett.},
  volume = {102},
  issue = {5},
  pages = {055301},
  numpages = {4},
  year = {2009},
  month = {Feb},
  publisher = {American Physical Society},
  doi = {10.1103/PhysRevLett.102.055301},
  url = {https://link.aps.org/doi/10.1103/PhysRevLett.102.055301}
}

@article{Bennett2000,
  title = {Exact and asymptotic measures of multipartite pure-state entanglement},
  author = {Bennett, Charles H. and Popescu, Sandu and Rohrlich, Daniel and Smolin, John A. and Thapliyal, Ashish V.},
  journal = {Phys. Rev. A},
  volume = {63},
  issue = {1},
  pages = {012307},
  numpages = {12},
  year = {2000},
  month = {Dec},
  publisher = {American Physical Society},
  doi = {10.1103/PhysRevA.63.012307},
  url = {https://link.aps.org/doi/10.1103/PhysRevA.63.012307}
}

@article{Eisert2001,
  title = {Schmidt measure as a tool for quantifying multiparticle entanglement},
  author = {Eisert, Jens and Briegel, Hans J.},
  journal = {Phys. Rev. A},
  volume = {64},
  issue = {2},
  pages = {022306},
  numpages = {4},
  year = {2001},
  month = {Jul},
  publisher = {American Physical Society},
  doi = {10.1103/PhysRevA.64.022306},
  url = {https://link.aps.org/doi/10.1103/PhysRevA.64.022306}
}

@article{Verstraete2003,
  title = {Normal forms and entanglement measures for multipartite quantum states},
  author = {Verstraete, Frank and Dehaene, Jeroen and De Moor, Bart},
  journal = {Phys. Rev. A},
  volume = {68},
  issue = {1},
  pages = {012103},
  numpages = {7},
  year = {2003},
  month = {Jul},
  publisher = {American Physical Society},
  doi = {10.1103/PhysRevA.68.012103},
  url = {https://link.aps.org/doi/10.1103/PhysRevA.68.012103}
}

@article{Ma2011,
  title = {Measure of genuine multipartite entanglement with computable lower bounds},
  author = {Ma, Zhi-Hao and Chen, Zhi-Hua and Chen, Jing-Ling and Spengler, Christoph and Gabriel, Andreas and Huber, Marcus},
  journal = {Phys. Rev. A},
  volume = {83},
  issue = {6},
  pages = {062325},
  numpages = {5},
  year = {2011},
  month = {Jun},
  publisher = {American Physical Society},
  doi = {10.1103/PhysRevA.83.062325},
  url = {https://link.aps.org/doi/10.1103/PhysRevA.83.062325}
}

@article{Hashemi2012,
  title = {Genuinely multipartite concurrence of $N$-qubit $X$ matrices},
  author = {Hashemi Rafsanjani, S. M. and Huber, M. and Broadbent, C. J. and Eberly, J. H.},
  journal = {Phys. Rev. A},
  volume = {86},
  issue = {6},
  pages = {062303},
  numpages = {6},
  year = {2012},
  month = {Dec},
  publisher = {American Physical Society},
  doi = {10.1103/PhysRevA.86.062303},
  url = {https://link.aps.org/doi/10.1103/PhysRevA.86.062303}
}

@article{Rieger_1994,
  title = {Zero-temperature quantum phase transition of a two-dimensional Ising spin glass},
  author = {Rieger, H. and Young, A. P.},
  journal = {Phys. Rev. Lett.},
  volume = {72},
  issue = {26},
  pages = {4141--4144},
  numpages = {0},
  year = {1994},
  month = {Jun},
  publisher = {American Physical Society},
  doi = {10.1103/PhysRevLett.72.4141},
  url = {https://link.aps.org/doi/10.1103/PhysRevLett.72.4141}
}

@article{Ahufinger_2005,
  title = {Disordered ultracold atomic gases in optical lattices: A case study of Fermi-Bose mixtures},
  author = {Ahufinger, V. and Sanchez-Palencia, L. and Kantian, A. and Sanpera, A. and Lewenstein, M.},
  journal = {Phys. Rev. A},
  volume = {72},
  issue = {6},
  pages = {063616},
  numpages = {25},
  year = {2005},
  month = {Dec},
  publisher = {American Physical Society},
  doi = {10.1103/PhysRevA.72.063616},
  url = {https://link.aps.org/doi/10.1103/PhysRevA.72.063616}
}

@article{Vojta2019,
  title = {Disorder in Quantum Many-Body Systems},
  volume = {10},
  ISSN = {1947-5462},
  url = {http://dx.doi.org/10.1146/annurev-conmatphys-031218-013433},
  DOI = {10.1146/annurev-conmatphys-031218-013433},
  number = {1},
  journal = {Annual Review of Condensed Matter Physics},
  publisher = {Annual Reviews},
  author = {Vojta,  Thomas},
  year = {2019},
  month = Mar,
  pages = {233–252}
}

@article{Shapiro_2012,
doi = {10.1088/1751-8113/45/14/143001},
url = {https://dx.doi.org/10.1088/1751-8113/45/14/143001},
year = {2012},
month = {mar},
publisher = {IOP Publishing},
volume = {45},
number = {14},
pages = {143001},
author = {Boris Shapiro},
title = {Cold atoms in the presence of disorder},
journal = {Journal of Physics A: Mathematical and Theoretical}
}

@article{Lewenstein2007,
  title = {Ultracold atomic gases in optical lattices: mimicking condensed matter physics and beyond},
  volume = {56},
  ISSN = {1460-6976},
  url = {http://dx.doi.org/10.1080/00018730701223200},
  DOI = {10.1080/00018730701223200},
  number = {2},
  journal = {Advances in Physics},
  publisher = {Informa UK Limited},
  author = {Lewenstein,  Maciej and Sanpera,  Anna and Ahufinger,  Veronica and Damski,  Bogdan and Sen(De),  Aditi and Sen,  Ujjwal},
  year = {2007},
  month = Mar,
  pages = {243–379}
}

@article{Mishra2016,
  title = {Constructive interference between disordered couplings enhances multiparty entanglement in quantum Heisenberg spin glass models},
  volume = {18},
  ISSN = {1367-2630},
  url = {http://dx.doi.org/10.1088/1367-2630/18/8/083044},
  DOI = {10.1088/1367-2630/18/8/083044},
  number = {8},
  journal = {New Journal of Physics},
  publisher = {IOP Publishing},
  author = {Mishra,  Utkarsh and Rakshit,  Debraj and Prabhu,  R and Sen(De),  Aditi and Sen,  Ujjwal},
  year = {2016},
  month = Aug,
  pages = {083044}
}

@article{Sadhukhan_2016,
  title = {Quantum correlations in quenched disordered spin models: Enhanced order from disorder by thermal fluctuations},
  author = {Sadhukhan, Debasis and Prabhu, R. and Sen(De), Aditi and Sen, Ujjwal},
  journal = {Phys. Rev. E},
  volume = {93},
  issue = {3},
  pages = {032115},
  numpages = {11},
  year = {2016},
  month = {Mar},
  publisher = {American Physical Society},
  doi = {10.1103/PhysRevE.93.032115},
  url = {https://link.aps.org/doi/10.1103/PhysRevE.93.032115}
}

@article{Aizenman_1989,
  title = {Rounding of first-order phase transitions in systems with quenched disorder},
  author = {Aizenman, Michael and Wehr, Jan},
  journal = {Phys. Rev. Lett.},
  volume = {62},
  issue = {21},
  pages = {2503--2506},
  numpages = {0},
  year = {1989},
  month = {May},
  publisher = {American Physical Society},
  doi = {10.1103/PhysRevLett.62.2503},
  url = {https://link.aps.org/doi/10.1103/PhysRevLett.62.2503}
}

@article{Niederberger_2008,
  title = {Disorder-Induced Order in Two-Component Bose-Einstein Condensates},
  author = {Niederberger, A. and Schulte, T. and Wehr, J. and Lewenstein, M. and Sanchez-Palencia, L. and Sacha, K.},
  journal = {Phys. Rev. Lett.},
  volume = {100},
  issue = {3},
  pages = {030403},
  numpages = {4},
  year = {2008},
  month = {Jan},
  publisher = {American Physical Society},
  doi = {10.1103/PhysRevLett.100.030403},
  url = {https://link.aps.org/doi/10.1103/PhysRevLett.100.030403}
}

@article{Niederberger2009,
  title = {Disorder-induced phase control in superfluid Fermi-Bose mixtures},
  volume = {86},
  ISSN = {1286-4854},
  url = {http://dx.doi.org/10.1209/0295-5075/86/26004},
  DOI = {10.1209/0295-5075/86/26004},
  number = {2},
  journal = {EPL (Europhysics Letters)},
  publisher = {IOP Publishing},
  author = {Niederberger,  A. and Wehr,  J. and Lewenstein,  M. and Sacha,  K.},
  year = {2009},
  month = Apr,
  pages = {26004}
}

@article{Niederberger_2010,
  title = {Disorder-induced order in quantum $\mathit{XY}$ chains},
  author = {Niederberger, A. and Rams, M. M. and Dziarmaga, J. and Cucchietti, F. M. and Wehr, J. and Lewenstein, M.},
  journal = {Phys. Rev. A},
  volume = {82},
  issue = {1},
  pages = {013630},
  numpages = {8},
  year = {2010},
  month = {Jul},
  publisher = {American Physical Society},
  doi = {10.1103/PhysRevA.82.013630},
  url = {https://link.aps.org/doi/10.1103/PhysRevA.82.013630}
}

@article{Bhattacharyya_2024,
  title = {Enhancing precision of atomic clocks by tuning disorder in accessories},
  author = {Bhattacharyya, Aparajita and Ghoshal, Ahana and Sen, Ujjwal},
  journal = {Phys. Rev. A},
  volume = {110},
  issue = {1},
  pages = {012620},
  numpages = {16},
  year = {2024},
  month = {Jul},
  publisher = {American Physical Society},
  doi = {10.1103/PhysRevA.110.012620},
  url = {https://link.aps.org/doi/10.1103/PhysRevA.110.012620}
}

@article{Pezze_2026,
  title = {Robust multipartite entanglement in dirty topological wires},
  author = {Pezz\`e, Luca and Lepori, Luca},
  journal = {Phys. Rev. B},
  volume = {113},
  issue = {20},
  pages = {205112},
  numpages = {15},
  year = {2026},
  month = {May},
  publisher = {American Physical Society},
  doi = {10.1103/n7k4-v7yw},
  url = {https://link.aps.org/doi/10.1103/n7k4-v7yw}
}

@article{Yousefjani2023,
  title = {Long-range interacting Stark many-body probes with super-Heisenberg precision},
  volume = {32},
  ISSN = {2058-3834},
  url = {http://dx.doi.org/10.1088/1674-1056/acf302},
  DOI = {10.1088/1674-1056/acf302},
  number = {10},
  journal = {Chinese Physics B},
  publisher = {IOP Publishing},
  author = {Yousefjani, Rozhin and He, Xingjian and Bayat, Abolfazl},
  year = {2023},
  month = Oct,
  pages = {100313}
}

@article{He_2023,
  title = {Stark Localization as a Resource for Weak-Field Sensing with Super-Heisenberg Precision},
  author = {He, Xingjian and Yousefjani, Rozhin and Bayat, Abolfazl},
  journal = {Phys. Rev. Lett.},
  volume = {131},
  issue = {1},
  pages = {010801},
  numpages = {7},
  year = {2023},
  month = {Jul},
  publisher = {American Physical Society},
  doi = {10.1103/PhysRevLett.131.010801},
  url = {https://link.aps.org/doi/10.1103/PhysRevLett.131.010801}
}

@article{Sahoo_2024,
  title = {Localization-driven quantum sensing},
  author = {Sahoo, Ayan and Mishra, Utkarsh and Rakshit, Debraj},
  journal = {Phys. Rev. A},
  volume = {109},
  issue = {3},
  pages = {L030601},
  numpages = {8},
  year = {2024},
  month = {Mar},
  publisher = {American Physical Society},
  doi = {10.1103/PhysRevA.109.L030601},
  url = {https://link.aps.org/doi/10.1103/PhysRevA.109.L030601}
}

@article{Sarkar_2025,
  title = {Noisy Stark probes as quantum-enhanced sensors},
  author = {Sarkar, Saubhik and Bayat, Abolfazl},
  journal = {Phys. Rev. A},
  volume = {111},
  issue = {6},
  pages = {062602},
  numpages = {11},
  year = {2025},
  month = {Jun},
  publisher = {American Physical Society},
  doi = {10.1103/PhysRevA.111.062602},
  url = {https://link.aps.org/doi/10.1103/PhysRevA.111.062602}
}

@article{Mothsara_2025,
  title = {Quantum-enhanced sensing with variable-range interactions},
  author = {Mothsara, Monika and Lakkaraju, Leela Ganesh Chandra and Ghosh, Srijon and Sen(De), Aditi},
  journal = {Phys. Rev. A},
  volume = {111},
  issue = {4},
  pages = {042628},
  numpages = {14},
  year = {2025},
  month = {Apr},
  publisher = {American Physical Society},
  doi = {10.1103/PhysRevA.111.042628},
  url = {https://link.aps.org/doi/10.1103/PhysRevA.111.042628}
}

@misc{Vishnupriya_2026,
      title={Robust quantum metrology using disordered probes}, 
      author={Vishnupriya K. and Harikrishnan K. J. and Amit Kumar Pal},
      year={2026},
      eprint={2604.11635},
      archivePrefix={arXiv},
      primaryClass={quant-ph},
      url={https://arxiv.org/abs/2604.11635}, 
}

@article{SanchezPalencia2010,
  title = {Disordered quantum gases under control},
  volume = {6},
  ISSN = {1745-2481},
  url = {http://dx.doi.org/10.1038/nphys1507},
  DOI = {10.1038/nphys1507},
  number = {2},
  journal = {Nature Physics},
  publisher = {Springer Science and Business Media LLC},
  author = {Sanchez-Palencia,  Laurent and Lewenstein,  Maciej},
  year = {2010},
  month = Feb,
  pages = {87–95}
}

@article{Bera_2016,
  title = {Disorder-induced enhancement and critical scaling of spontaneous magnetization in random-field quantum spin systems},
  author = {Bera, Anindita and Rakshit, Debraj and Lewenstein, Maciej and Sen(De), Aditi and Sen, Ujjwal and Wehr, Jan},
  journal = {Phys. Rev. B},
  volume = {94},
  issue = {1},
  pages = {014421},
  numpages = {12},
  year = {2016},
  month = {Jul},
  publisher = {American Physical Society},
  doi = {10.1103/PhysRevB.94.014421},
  url = {https://link.aps.org/doi/10.1103/PhysRevB.94.014421}
}

@article{Karimipour_2013,
  title = {Kitaev-Ising model and the transition between topological and ferromagnetic order},
  author = {Karimipour, Vahid and Memarzadeh, Laleh and Zarkeshian, Parisa},
  journal = {Phys. Rev. A},
  volume = {87},
  issue = {3},
  pages = {032322},
  numpages = {9},
  year = {2013},
  month = {Mar},
  publisher = {American Physical Society},
  doi = {10.1103/PhysRevA.87.032322},
  url = {https://link.aps.org/doi/10.1103/PhysRevA.87.032322}
}

@article{Halasz_2012,
  title = {Probing topological order with R\'enyi entropy},
  author = {Hal\'asz, G\'abor B. and Hamma, Alioscia},
  journal = {Phys. Rev. A},
  volume = {86},
  issue = {6},
  pages = {062330},
  numpages = {12},
  year = {2012},
  month = {Dec},
  publisher = {American Physical Society},
  doi = {10.1103/PhysRevA.86.062330},
  url = {https://link.aps.org/doi/10.1103/PhysRevA.86.062330}
}

@article{Tsomokos09,
  title = {Topological order following a quantum quench},
  author = {Tsomokos, Dimitris I. and Hamma, Alioscia and Zhang, Wen and Haas, Stephan and Fazio, Rosario},
  journal = {Phys. Rev. A},
  volume = {80},
  issue = {6},
  pages = {060302(R)},
  numpages = {4},
  year = {2009},
  month = {Dec},
  publisher = {American Physical Society},
  doi = {10.1103/PhysRevA.80.060302},
  url = {https://link.aps.org/doi/10.1103/PhysRevA.80.060302}
}

@article{Zeng16,
  title = {Thermalization of topological entropy after a quantum quench},
  author = {Zeng, Yu and Hamma, Alioscia and Fan, Heng},
  journal = {Phys. Rev. B},
  volume = {94},
  issue = {12},
  pages = {125104},
  numpages = {20},
  year = {2016},
  month = {Sep},
  publisher = {American Physical Society},
  doi = {10.1103/PhysRevB.94.125104},
  url = {https://link.aps.org/doi/10.1103/PhysRevB.94.125104}
}

@article{HalaszHamma13,
  title = {Topological R\'enyi Entropy after a Quantum Quench},
  author = {Hal\'asz, G\'abor B. and Hamma, Alioscia},
  journal = {Phys. Rev. Lett.},
  volume = {110},
  issue = {17},
  pages = {170605},
  numpages = {5},
  year = {2013},
  month = {Apr},
  publisher = {American Physical Society},
  doi = {10.1103/PhysRevLett.110.170605},
  url = {https://link.aps.org/doi/10.1103/PhysRevLett.110.170605}
}

@article{Wen2002,
  title = {Quantum orders and symmetric spin liquids},
  author = {Wen, Xiao-Gang},
  journal = {Phys. Rev. B},
  volume = {65},
  issue = {16},
  pages = {165113},
  numpages = {37},
  year = {2002},
  month = {Apr},
  publisher = {American Physical Society},
  doi = {10.1103/PhysRevB.65.165113},
  url = {https://link.aps.org/doi/10.1103/PhysRevB.65.165113}
}

@article{Wen1995,
author = {Xiao-Gang Wen},
title = {Topological orders and edge excitations in fractional quantum Hall states},
journal = {Advances in Physics},
volume = {44},
number = {5},
pages = {405--473},
year = {1995},
publisher = {Taylor \& Francis},
doi = {10.1080/00018739500101566},
URL = { 
        https://doi.org/10.1080/00018739500101566
}
}

@article{Wen2006,
  title = {Detecting Topological Order in a Ground State Wave Function},
  author = {Levin, Michael and Wen, Xiao-Gang},
  journal = {Phys. Rev. Lett.},
  volume = {96},
  issue = {11},
  pages = {110405},
  numpages = {4},
  year = {2006},
  month = {Mar},
  publisher = {American Physical Society},
  doi = {10.1103/PhysRevLett.96.110405},
  url = {https://link.aps.org/doi/10.1103/PhysRevLett.96.110405}
}

@article{kitaev_preskill_2006,
  title = {Topological Entanglement Entropy},
  author = {Kitaev, Alexei and Preskill, John},
  journal = {Phys. Rev. Lett.},
  volume = {96},
  issue = {11},
  pages = {110404},
  numpages = {4},
  year = {2006},
  month = {Mar},
  publisher = {American Physical Society},
  doi = {10.1103/PhysRevLett.96.110404},
  url = {https://link.aps.org/doi/10.1103/PhysRevLett.96.110404}
}

@article{kitaev_preskill_2002,
    author = {Dennis, Eric and Kitaev, Alexei and Landahl, Andrew and Preskill, John},
    title = {Topological quantum memory},
    journal = {Journal of Mathematical Physics},
    volume = {43},
    number = {9},
    pages = {4452-4505},
    year = {2002},
    month = {09},
    issn = {0022-2488},
    doi = {10.1063/1.1499754},
    url = {https://doi.org/10.1063/1.1499754}
}

@article{PezzeRMP18,
  author = {Pezz{\`e}, Luca and Smerzi, Augusto and Oberthaler, Markus K. and Schmied, Roman and Treutlein, Philipp},
  title = {Quantum metrology with nonclassical states of atomic ensembles},
  journal = {Rev. Mod. Phys.},
  volume = {90},
  pages = {035005},
  year = {2018},
  doi = {10.1103/RevModPhys.90.035005},
  url = {https://doi.org/10.1103/RevModPhys.90.035005}
}

@article{IeminiTimeCrystal24,
  author = {Iemini, Fernando and Fazio, Rosario and Sanpera, Anna},
  title = {Floquet time crystals as quantum sensors of ac fields},
  journal = {Phys. Rev. A},
  volume = {109},
  pages = {L050203},
  year = {2024},
  doi = {10.1103/PhysRevA.109.L050203},
  url = {https://doi.org/10.1103/PhysRevA.109.L050203}
}

@article{LakkarajuToricDynamics24,
  author = {Lakkaraju, Leela Ganesh Chandra and Haldar, Sudip Kumar and Sen(De), Aditi},
  title = {Predicting a topological quantum phase transition from dynamics via multisite entanglement},
  journal = {Phys. Rev. A},
  volume = {109},
  pages = {022436},
  year = {2024},
  doi = {10.1103/PhysRevA.109.022436},
  url = {https://doi.org/10.1103/PhysRevA.109.022436}
}

@article{BoixoDynamics08,
  author = {Boixo, Sergio and Datta, Animesh and Davis, Matthew J. and Flammia, Steven T. and Shaji, Anil and Caves, Carlton M.},
  title = {Quantum Metrology: Dynamics versus Entanglement},
  journal = {Phys. Rev. Lett.},
  volume = {101},
  pages = {040403},
  year = {2008},
  doi = {10.1103/PhysRevLett.101.040403},
  url = {https://doi.org/10.1103/PhysRevLett.101.040403}
}

@article{Zhang_2022,
  title = {Multipartite entanglement of the topologically ordered state in a perturbed toric code},
  author = {Zhang, Yu-Ran and Zeng, Yu and Liu, Tao and Fan, Heng and You, J. Q. and Nori, Franco},
  journal = {Phys. Rev. Res.},
  volume = {4},
  issue = {2},
  pages = {023144},
  numpages = {11},
  year = {2022},
  month = {May},
  publisher = {American Physical Society},
  doi = {10.1103/PhysRevResearch.4.023144},
  url = {https://link.aps.org/doi/10.1103/PhysRevResearch.4.023144}
}

@article{TrebstToric07,
  author = {Trebst, Simon and Werner, Philipp and Troyer, Matthias and Shtengel, Kirill and Nayak, Chetan},
  title = {Breakdown of a Topological Phase: Quantum Phase Transition in a Loop Gas Model with Tension},
  journal = {Phys. Rev. Lett.},
  volume = {98},
  pages = {070602},
  year = {2007},
  doi = {10.1103/PhysRevLett.98.070602},
  url = {https://doi.org/10.1103/PhysRevLett.98.070602}
}

@misc{Preskill1998,
  title = {Lecture Notes for Physics 229: Quantum Information and Computation},
  author = {John Preskill},
  year = {1998},
  institution = {California Institute of Technology},
  url = {https://www.preskill.caltech.edu/ph229/}
}

@misc{lidar_2020_lecture,
      title={Lecture Notes on the Theory of Open Quantum Systems}, 
      author={Daniel A. Lidar},
      year={2020},
      eprint={1902.00967},
      archivePrefix={arXiv},
      primaryClass={quant-ph}
}

@book{open_quan_book,
author = {Breuer, Heinz-Peter and Petruccione, Francesco},
    title = {The Theory of Open Quantum Systems},
    publisher = {Oxford University Press},
    year = {2007},
    month = {01},
    isbn = {9780199213900},
    doi = {10.1093/acprof:oso/9780199213900.001.0001},
    url = {https://doi.org/10.1093/acprof:oso/9780199213900.001.0001},
}

@book{Rivas2012,
  title = {Open Quantum Systems: An Introduction},
  ISBN = {9783642233548},
  ISSN = {2191-5431},
  url = {http://dx.doi.org/10.1007/978-3-642-23354-8},
  DOI = {10.1007/978-3-642-23354-8},
  journal = {SpringerBriefs in Physics},
  publisher = {Springer Berlin Heidelberg},
  author = {Rivas,  Angel and Huelga,  Susana F.},
  year = {2012}
}

@article{Zurek2003,
    author  = {Zurek, Wojciech Hubert},
    title   = {Decoherence, einselection, and the quantum origins of the classical},
    journal = {Reviews of Modern Physics},
    volume  = {75},
    pages   = {715--775},
    year    = {2003},
    doi     = {10.1103/RevModPhys.75.715}
  }

@article{Sen_2006,
  title = {Quantum-information processing in disordered and complex quantum systems},
  author = {Sen(De), Aditi and Sen, Ujjwal and Ahufinger, Veronica and Briegel, Hans J. and Sanpera, Anna and Lewenstein, Maciej},
  journal = {Phys. Rev. A},
  volume = {74},
  issue = {6},
  pages = {062309},
  numpages = {8},
  year = {2006},
  month = {Dec},
  publisher = {American Physical Society},
  doi = {10.1103/PhysRevA.74.062309},
  url = {https://link.aps.org/doi/10.1103/PhysRevA.74.062309}
}

@article{Sherrington_1975,
  title = {Solvable Model of a Spin-Glass},
  author = {Sherrington, David and Kirkpatrick, Scott},
  journal = {Phys. Rev. Lett.},
  volume = {35},
  issue = {26},
  pages = {1792--1796},
  numpages = {0},
  year = {1975},
  month = {Dec},
  publisher = {American Physical Society},
  doi = {10.1103/PhysRevLett.35.1792},
  url = {https://link.aps.org/doi/10.1103/PhysRevLett.35.1792}
}

@article{Edwards1975,
  title = {Theory of spin glasses},
  volume = {5},
  ISSN = {0305-4608},
  url = {http://dx.doi.org/10.1088/0305-4608/5/5/017},
  DOI = {10.1088/0305-4608/5/5/017},
  number = {5},
  journal = {Journal of Physics F: Metal Physics},
  publisher = {IOP Publishing},
  author = {Edwards,  S F and Anderson,  P W},
  year = {1975},
  month = May,
  pages = {965–974}
}

@article{Derrida_1980,
  title = {Random-Energy Model: Limit of a Family of Disordered Models},
  author = {Derrida, B.},
  journal = {Phys. Rev. Lett.},
  volume = {45},
  issue = {2},
  pages = {79--82},
  numpages = {0},
  year = {1980},
  month = {Jul},
  publisher = {American Physical Society},
  doi = {10.1103/PhysRevLett.45.79},
  url = {https://link.aps.org/doi/10.1103/PhysRevLett.45.79}
}

@article{Wasilewski_2010,
  title = {Quantum Noise Limited and Entanglement-Assisted Magnetometry},
  author = {Wasilewski, W. and Jensen, K. and Krauter, H. and Renema, J. J. and Balabas, M. V. and Polzik, E. S.},
  journal = {Phys. Rev. Lett.},
  volume = {104},
  issue = {13},
  pages = {133601},
  numpages = {4},
  year = {2010},
  month = {Mar},
  publisher = {American Physical Society},
  doi = {10.1103/PhysRevLett.104.133601},
  url = {https://link.aps.org/doi/10.1103/PhysRevLett.104.133601}
}

@article{Escher2011,
  title = {General framework for estimating the ultimate precision limit in noisy quantum-enhanced metrology},
  volume = {7},
  ISSN = {1745-2481},
  url = {http://dx.doi.org/10.1038/nphys1958},
  DOI = {10.1038/nphys1958},
  number = {5},
  journal = {Nature Physics},
  publisher = {Springer Science and Business Media LLC},
  author = {Escher,  B. M. and de Matos Filho,  R. L. and Davidovich,  L.},
  year = {2011},
  month = Mar,
  pages = {406–411}
}

@article{DemkowiczDobrzaski2012,
  title = {The elusive Heisenberg limit in quantum-enhanced metrology},
  volume = {3},
  ISSN = {2041-1723},
  url = {http://dx.doi.org/10.1038/ncomms2067},
  DOI = {10.1038/ncomms2067},
  number = {1},
  journal = {Nature Communications},
  publisher = {Springer Science and Business Media LLC},
  author = {Demkowicz-Dobrzański,  Rafał and Kołodyński,  Jan and Guţă,  Mădălin},
  year = {2012},
  month = {Sept} 
}

@article{Koodyski2013,
  title = {Efficient tools for quantum metrology with uncorrelated noise},
  volume = {15},
  ISSN = {1367-2630},
  url = {http://dx.doi.org/10.1088/1367-2630/15/7/073043},
  DOI = {10.1088/1367-2630/15/7/073043},
  number = {7},
  journal = {New Journal of Physics},
  publisher = {IOP Publishing},
  author = {Kołodyński,  Jan and Demkowicz-Dobrzański,  Rafał},
  year = {2013},
  month = {July},
  pages = {073043}
}

@article{Alipour_2014,
  title = {Quantum Metrology in Open Systems: Dissipative Cram\'er-Rao Bound},
  author = {Alipour, S. and Mehboudi, M. and Rezakhani, A. T.},
  journal = {Phys. Rev. Lett.},
  volume = {112},
  issue = {12},
  pages = {120405},
  numpages = {6},
  year = {2014},
  month = {Mar},
  publisher = {American Physical Society},
  doi = {10.1103/PhysRevLett.112.120405},
  url = {https://link.aps.org/doi/10.1103/PhysRevLett.112.120405}
}

@article{haase2016_open_review,
  title={Precision limits in quantum metrology with open quantum systems},
  author={Haase, Jan F and Smirne, Andrea and Huelga, SF and Ko{\l}odynski, J and Demkowicz-Dobrzanski, R},
  journal={Quantum Measurements and Quantum Metrology},
  volume={5},
  number={1},
  pages={13--39},
  year={2016},
  publisher={De Gruyter Open Access}
}

@article{Falaye2017,
  title = {Investigating quantum metrology in noisy channels},
  volume = {7},
  ISSN = {2045-2322},
  url = {http://dx.doi.org/10.1038/s41598-017-16710-w},
  DOI = {10.1038/s41598-017-16710-w},
  number = {1},
  journal = {Scientific Reports},
  publisher = {Springer Science and Business Media LLC},
  author = {Falaye,  B. J. and Adepoju,  A. G. and Aliyu,  A. S. and Melchor,  M. M. and Liman,  M. S. and Oluwadare,  O. J. and González-Ramírez,  M. D. and Oyewumi,  K. J.},
  year = {2017},
  month = Nov 
}

\appendix

\section{Multipartite entanglement}
\label{app:ggm_definition}

An \(N\)-party pure state $| \Psi_{1,2,\ldots,N}\rangle$ is genuinely multipartite entangled (GME), when it is entangled in all bipartitions $\mathfrak{B}=\{(A:B) \mid A\cap B\!=\!\emptyset, \lVert A\cup B\rVert\!=\!N\}$ of the system. The GME of the pure state can be characterized by its distance from the set of non-genuinely (nG) entangled states, i.e., the set of states which are separable at some bipartition $A:B$. This provides a distance-based measure, known as the generalized geometric measure (GGM)~\cite{Wei2003, Sen2010, ggm_shimony, Barnum_ggm_2}, and is defined as
\begin{align}
    \mathcal{G}(|\Psi\rangle) =& 1-\underset{|\upsilon\rangle \in \text{nG}}{\max}|\langle\upsilon|\Psi\rangle|^2 = 1-\max_{(A:B)\in\mathfrak{B}}[\xi_{A:B}^2]
    \label{eq:ggm_full}
\end{align}
where the maximization is over all bipartitions \((A:B)\), and \(\xi_{A:B}\) is the largest Schmidt coefficient in the corresponding bipartition. As the number of possible bipartitions increases exponentially with the system size $N$, we simplify the computation of GGM by restricting the maximization to single-site bipartitions, i.e., with $\lVert A\rVert=1$. This simplification defines the geometric measure (GM) as
\begin{align}
    \mathcal{G}_1(| \Psi\rangle)=1-\max_{(A:B)\in\mathfrak{B}}[\xi_{A:B}^2 \mid \lVert A\rVert=1].
\end{align}
The validity of approximation is seen with 
$\Delta\mathcal{G} = \mathcal{G}_{1} - \mathcal{G} = 0$ due to the diagonal nature of the subsystem's density matrix \cite{LakkarajuToricDynamics24}, and use $\mathcal{G}_{1}$ and $\mathcal{G}$ interchangeably in the work.

\end{document}